# Confinement Controlled Phase-Pure 2D Layered Perovskites for Band-Edge Engineering

*Danila A. Tatarinov*[1,2], *Alexander Schleusener*[1], *and Roman Krahne*[1]

[1]*Optoelectronics, Italian Institute of Technology (IIT), Via Morego 30, 16163 Genoa, Italy*
[2]*Dipartimento di Chimica e Chimica Industriale, Università degli Studi di Genova, Via Dodecaneso, 31, Genova 16146, Italy*

*Corresponding authors email:roman.krahne@iit.it*

Funding: HE-MSCA-SE project "DELIGHT", grant number 101131111

Abstract

Controlling the number *n* of connected octahedra layers in two-dimensional metal halide perovskites enables fine control of the semiconductor band gap that is highly advantageous for optoelectronic and light emission applications. Phase purity in terms of *n* is crucial towards narrow emission bands and the design of band structure alignment for charge carrier funneling. We present a simple and scalable synthesis method using cesium and butylammonium as A-cations to fabricate microcrystal powders of 2D lead bromide perovskites that results in samples with very high *n*-phase purity. The obtained microcrystals show a sharp onset in absorbance and single narrow emission peak in UV-blue spectral range, and we confirm their structure and composition with X-ray diffraction and energy dispersive spectroscopy. We show that we can tailor the spectral absorption of microcrystals samples by controlled mixing of the *n*-phase pure powders, and we demonstrate this capability with spectral photocurrent measurements using pellets pressed from pure and mixed powders. Both the spectral and dynamic response of the photocurrent of the pellets depend strongly on the composition of the powders, highlighting the design potential of our approach towards band-edge engineering.

# 1. Introduction

Two-dimensional layered metal-halide perovskites (2DLPs) have emerged as a highly versatile class of solution-processable semiconductors for applications ranging from photovoltaics and light emission to photodetection and photonics.[1–7] In these materials, semiconducting metal–halide octahedral layers are separated by bulky organic spacer cations, giving rise to strong excitonic effects, high refractive indices, and pronounced quantum and dielectric confinement.[8–11] Introducing small A-site cations such as $Cs^+$, methylammonium ($MA^+$), or formamidinium ($FA^+$) increases the number of the $[PbX_6]^{4-}$ octahedral layers (*n*), thereby reducing quantum confinement within the inorganic slabs. As illustrated in Scheme 1, increasing the number of inorganic layers (*n*) leads to a progressive reduction of the band gap and a transition from strongly confined to more delocalized electronic states.[12–16] This provides a versatile means to tune the band gap, exciton diffusion, and charge-carrier mobility in these semiconductors.[17–21]

However, the synthesis of phase-pure 2DLPs with $n > 1$ remains challenging due to the similar formation energies of different *n* members within a given 2D perovskite material system.[22–25] As a result, multiple *n* phases often coexist, leading to broadened optical features, increased energetic disorder, and limited reproducibility in device performance.[26–29] Achieving phase-pure 2D layered perovskites is therefore particularly demanding, especially when scalable processing routes are desired.[3,12,26,30,31]

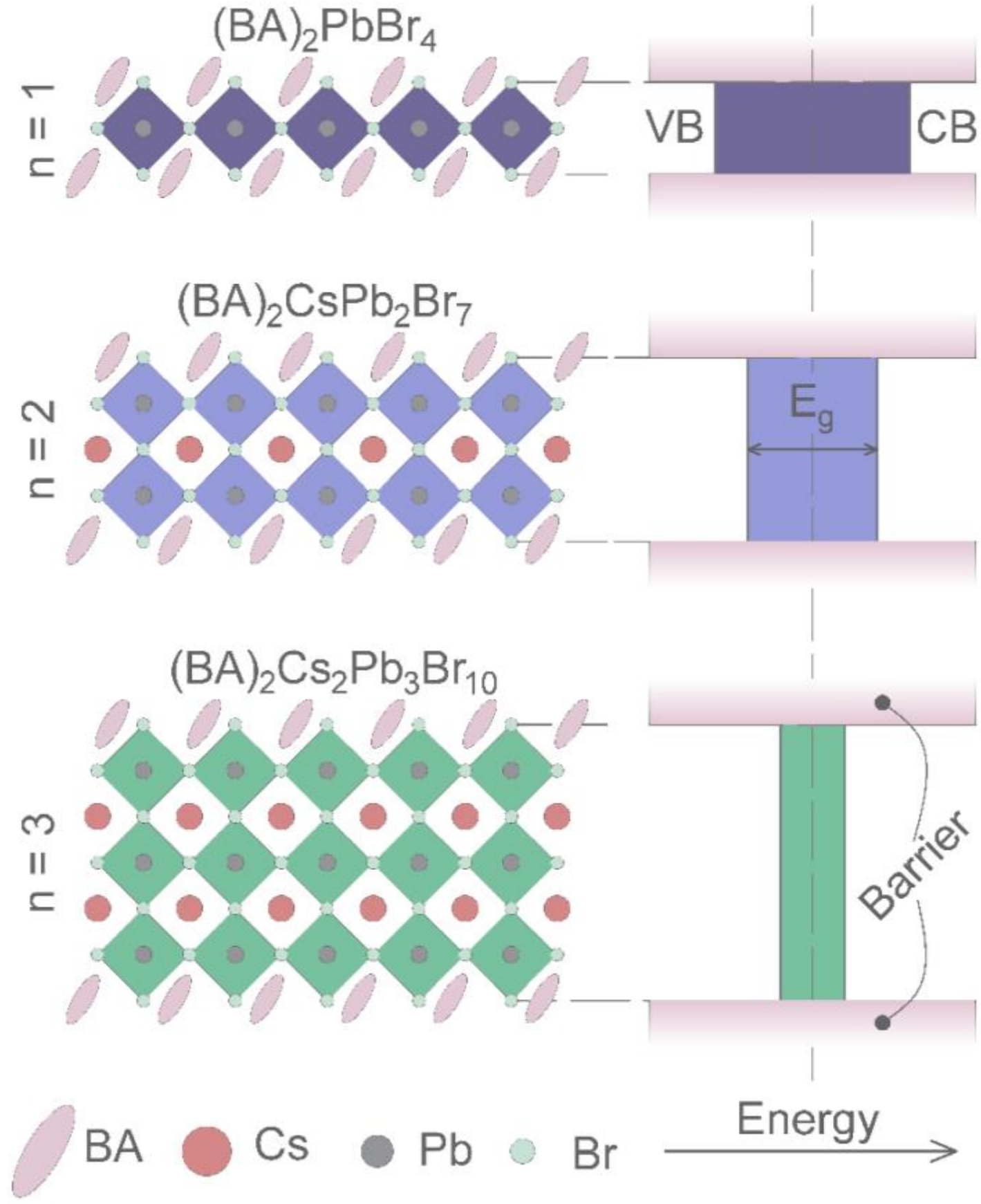

**Scheme 1**. Structural representation of phase-pure layered perovskites with different inorganic layer thicknesses (*n* = 1–3) and their corresponding band structures. Increasing *n* leads to a reduction of the bandgap ($E_g$), reflecting the transition from strongly confined to more extended electronic states.

Recently, progressive *n*-phase transformation has been demonstrated in MA-based layered perovskites using kinetically controlled crystallization strategies.[32] In these systems, phase evolution is enabled by the mobility and intercalation of MA cations, together with iodide and Pb species, allowing *n* to increase with crystallization time or temperature. This results in a continuous evolution from lower to higher *n*-phases during growth. While this approach provided important mechanistic insights, its applicability to other halides and to inorganic A-site cations remains largely unexplored. Other approaches rely on controlling the crystallization environment, for example through co-solvent engineering, where phase purity is achieved by suppressing supersaturation and uncontrolled nucleation during growth.[19,33,34] While such methods effectively improve crystallinity and reduce phase inhomogeneity, they primarily focus on controlling nucleation pathways.

In this work, we show that extended dwell times promote phase stabilization without the need for additional solvent engineering. To this aim, we investigate the phase evolution in Cs-containing 2D bromide perovskites of the formula $BA_2Cs_{n-1}Pb_nBr_{3n+1}$ (BA, butylammonium) and demonstrate a distinct kinetic pathway toward phase purification. Using an acid-based synthesis at elevated temperature followed by controlled cooling, we show that the final phase composition is governed primarily by the isothermal dwell time rather than the cooling rate. *In-situ* photoluminescence spectroscopy directly reveals the transformation from mixed-phase ensembles to phase-pure $n = 2$ ($BA_2CsPb_2Br_7$) and $n = 3$ ($BA_2Cs_2Pb_3Br_{10}$) materials, highlighting a phase-selection mechanism that does not rely on A-site cation intercalation. Furthermore, in many applications a more flexible design of the band gap absorption is desirable that could be obtained by a controlled overlap of the absorption features of the different *n*-phases.[35,36] We address this challenge by a controlled phase mixing strategy, in which phase-pure layered perovskites are first synthesized and subsequently combined in defined ratios to form mixed-phase systems.

We characterize the charge transport and photoconductivity of the different *n*-phases using pellets pressed from the respective powders, and we demonstrate that controlled phase mixing enables tuning of spectral photocurrent profiles and results in faster photoresponse. Our results establish controlled phase mixing as a simple and scalable post-synthetic strategy to engineer both band-edge properties and charge transport in layered perovskites, providing a new route toward tunable optoelectronic materials and devices.

## 2. Results and Discussion

### 2.1. Formation of phase-pure 2DLP powders

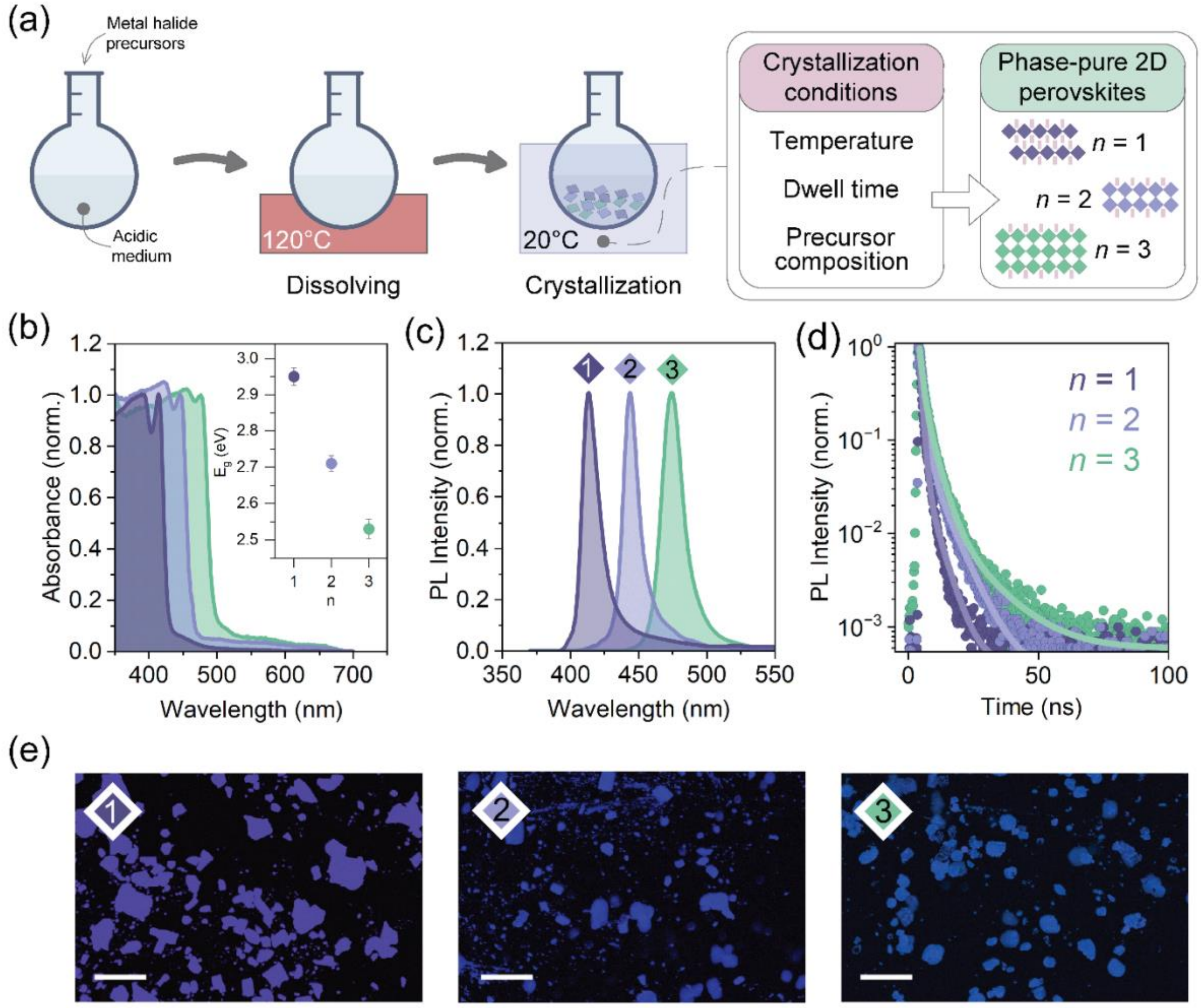

**Figure 1. Phase-selective crystallization and optical characterization of phase-pure 2D layered perovskite microcrystals.** (a) Schematic illustration of the acid-assisted synthesis strategy involving dissolution of precursor salts at 120 °C followed by controlled crystallization, where crystallization conditions govern the selective formation of $n$ = 1–3 layered perovskite phases. (b) Normalized absorption spectra of the phase-pure $n$ = 1–3 samples, the inset shows the corresponding optical band gaps. (c) Steady-state PL spectra ($\lambda_{ex}$ = 365 nm). (d) Time-resolved PL decay curves with multi-exponential fits ($\lambda_{ex}$ = 371 nm). (e) Confocal fluorescence microscopy (scale bars: 200 µm) of the synthesized microcrystals ($\lambda_{ex}$ = 405 nm).

Figure 1 summarizes the synthesis and optical characterization of phase-pure 2DLP microcrystals with different $n$ phases. The synthesis procedure is illustrated in Fig. 1a and consists of two main steps: dissolution of the precursor salts in an acidic solution at 120 °C followed by controlled crystallization in a water bath at 20 °C over several hours. The resulting microcrystals were subsequently isolated and dried. We first address the properties of the obtained materials, and after that discuss their growth process.

The synthesized powders are homogeneous and exhibit characteristic colors, ranging from white for $n = 1$ to bright yellow and orange for $n = 2$ and $n = 3$, respectively. All samples display bright photoluminescence (PL) under UV excitation (Figure S1). The proposed synthetic protocol can be readily scaled up by at least an order of magnitude without modification of the procedure, enabling the preparation of gram-scale quantities of phase-pure products (Figure S2). The absorption and PL spectra recorded from the final products are displayed in Fig. 1b,c. The absorption spectra (Figure 1b) exhibit well-defined excitonic features for all compositions ($n$ = 1–3), with a clear shift of the absorption edge as a function of $n$. The corresponding optical band gap values, extracted from Tauc plots (see Figure S3), are summarized in the inset of Figure 1b. Due to the increased quantum confinement, a progressive blue shift of both the absorption edge and PL maximum is observed with decreasing $n$.[37,38] Time-resolved PL measurements (Figure 1d) reveal a gradual increase in the average PL decay lifetime with increasing $n$ value, as evidenced by multi-exponential fitting of the decay traces (Figure S4 and Table S1) , consistent with previous reports.[39] Furthermore, representative PL confocal images (Figure 1e) for the $n = 2$ and $n = 3$ phases show a homogeneous PL signal across the scanned area, where the uniform optical response underpins the phase purity.

We investigated the structural and compositional properties of the powders with different $n$ phase by X-ray diffraction (XRD) and energy dispersive spectroscopy (EDS), and show the results in Figures S5–S7. The experimental XRD patterns are in agreement with the simulated ones reported [39] for the corresponding 2DLP phases, while no additional diffraction peaks associated with secondary phases or higher-dimensional perovskites are observed. EDS elemental mapping further confirms the homogeneous spatial distribution of Cs, Pb, and Br throughout the individual microcrystals, while the measured elemental compositions are in good agreement with the expected stoichiometry for each *n* phase.

To gain deeper insight into the growth kinetics, we investigated the synthesis under a wide range of cooling conditions by monitoring the temperature (see Figures S8–S9 and Table S2) and PL of the growth solutions throughout the crystallization process (Figure 2). The reaction is initiated at a solution temperature of 120 °C, which is required to fully dissolve the precursors. Under these conditions, the solution is saturated with respect to the perovskite precursors, providing the driving force for crystallization upon cooling. The reaction vial is then transferred to water baths at different temperatures. The water bath proved to be the most versatile approach since it allowed *in-situ* monitoring of the PL while maintaining a constant temperature (after the relatively fast initial cooling upon transfer) and minimizing environmental fluctuations, resulting in phase-pure powders within a reasonable reaction time. We tested our synthesis protocol with other cooling approaches converging to different temperatures, namely with an ice bath and in a cold room environment (see Figure S10). In both cases, after 12 h of reaction time, still different *n*-phases were present (probed by the PL), however, we cannot exclude that after much longer times also with these approaches phase pure materials could be obtained. Due to the relatively fast cooling in the water bath (with an exponential temperature decrease), the time to reach a specified temperature is much shorter than the overall growth time. We therefore conclude that the set temperature and time after cooling are the key parameters governing the process to stabilize a single *n*-phase.

To verify this hypothesis, we systematically investigated crystallization over a broad temperature range (3–60 °C). In-situ PL measurements obtained under different crystallization temperatures (Figures S11 and S12) show that phase-selective crystallization occurs only within a relatively narrow temperature window. At low temperatures (3–15 °C), crystallization proceeds slowly and multiple emissive phases coexist for extended periods, resulting in mixed-phase products even after prolonged reaction times. In contrast, at 20 °C, PL features from several *n*-phases are initially observed during the first 15 min (Figure 2a–d), indicating the coexistence of multiple phases after the solution temperature has stabilized. As crystallization proceeds, these intermediate phases gradually disappear, yielding a single stable PL peak characteristic of the target phase. Phase-pure microcrystals are obtained after 5 h for the $n = 2$ composition and after 10 h for the $n = 3$ composition.

The phase maps (Figures S13 and S14) and XRD analysis (Figure S15) support these observations. For both $n = 2$ and $n = 3$ compositions, phase-pure products are reproducibly obtained only near 20 °C. Samples synthesized at 3 °C exhibit diffraction patterns characteristic of mixed layered phases, whereas those prepared at 45 °C show diffraction patterns inconsistent with the target layered perovskite structures, confirming the formation of competing non-perovskite phases.

Overall, these results demonstrate that phase-selective crystallization is governed primarily by the final crystallization temperature and the growth time at this temperature, which together control the transformation of the initially formed phases. At lower temperatures, phase transformation is kinetically hindered, resulting in the prolonged coexistence of metastable phases. In contrast, at higher temperatures, competing non-perovskite crystallization pathways become dominant. Only at an intermediate temperature of approximately 20 °C are the kinetic conditions favorable for the gradual transformation of these intermediate phases into the targeted phase, consistent with the crystallization pathway proposed in Figure 2e.

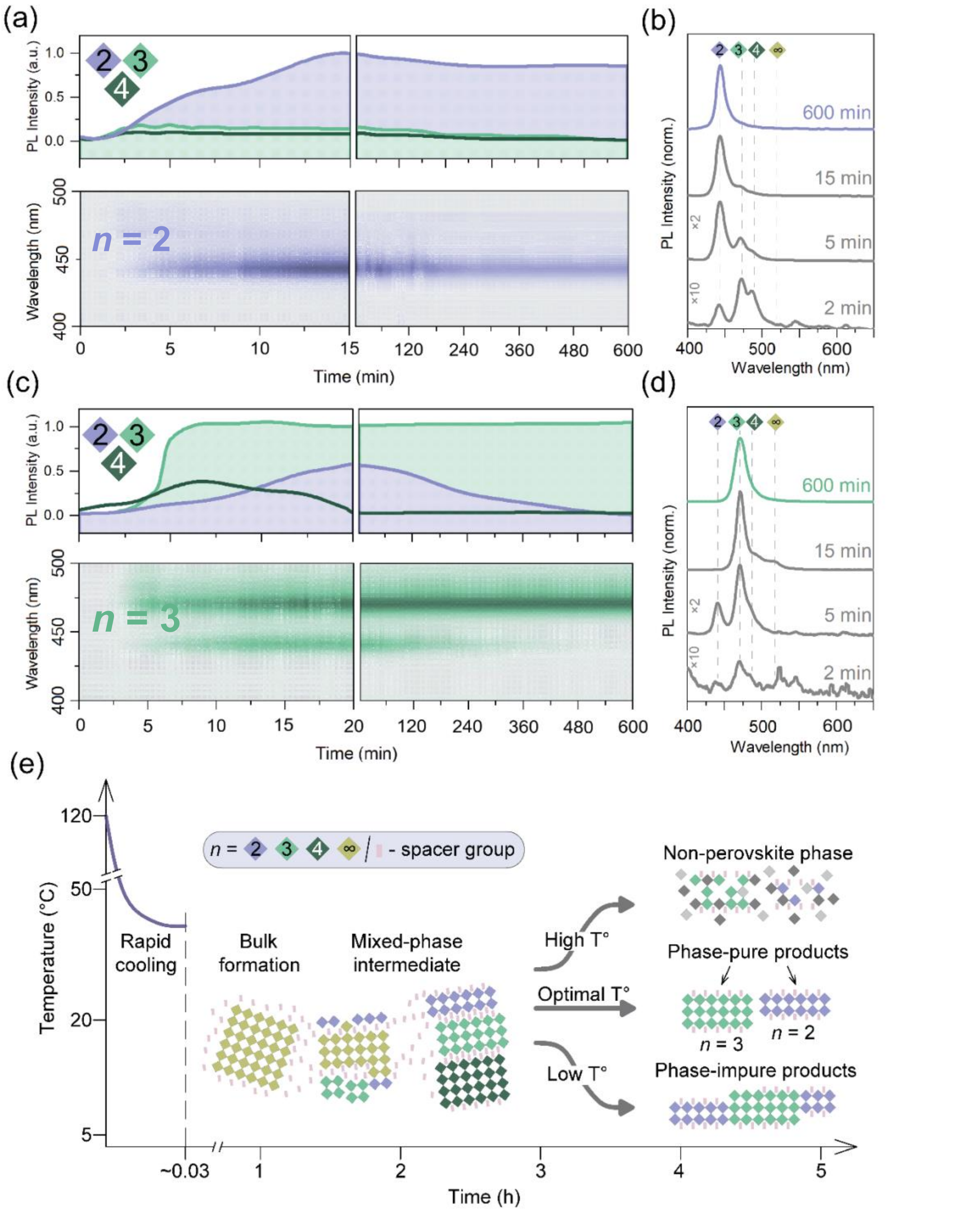

**Figure 2.** PL evolution versus time during the reaction process of 2D layered perovskites. (a,c) Normalized PL intensity profiles (top) and corresponding PL color maps (bottom) as a function of time, showing the spectral evolution over short (0–15 min) and extended (15–600 min) time scales for the growth process aiming at $n = 2$ (a) and $n = 3$ (c) samples, respectively. (b,d)

Representative normalized PL spectra recorded at selected crystallization times illustrating the evolution of the spectral emission features. (e) Schematic illustration of the proposed crystallization pathway leading to the selective formation of phase-pure 2D layered perovskites.

## 2.2. Phase-pure and mixed 2DLP powders for band-edge engineering

To demonstrate the band-edge engineering we pressed the powders into thin pellets that can be easily contacted and handled, and we investigate their (photo)electrical properties. The pellets were pressed at a pressure of 200 MPa, and their dimensions were 5 mm diameter (see Figure 3, insets) and around 200 µm height (Fig. S16-S18). For the back contact the pellets were placed on a conductive carbon tape, and the top contacts were fabricated by sputtering of 50 nm Au using a shadow mask.

We first address the electrical and photoconductive properties of pellets fabricated from the phase-pure powders. We find that the $n$ = 1 phase pellets show a much lower dark current compared to the $n$ = 2 and n = 3 phases, with nearly ohmic behavior (~40 GΩ) and negligible hysteresis. Upon illumination at a bias of 5V, we observe a fast photoresponse (Figure 3c) with rise and fall times on a timescale of tens of milliseconds, and low responsivity below 1 mA/W. The (photo)electrical behavior is very different for the two higher $n$ materials that show 3–4 orders higher dark current with marked hysteresis, and a strong, but slow photoresponse. The hysteretic behavior is reproducible over several consecutive scans, which indicates stable ionic current pathways. Interestingly, the hysteresis features a shape that is more open (extended) around zero voltage bias, and narrows towards higher bias voltages for both polarities. This points to ionic currents mediated by shallow traps that can be effective under small electric fields. We speculate that the conductive pathways in the pellets are dominated by the edges of the microcrystals, and that halide migration is affecting them. Halide migration has been reported to be more dominant in higher-*n* 2D layered perovskites due to the higher density of the halides in the bulk phase.[40]

In light-on/light-off measurements, the current time traces of the $n = 2$ and $n = 3$ phase pellets show a characteristic saw-tooth pattern that originates from very slow rise and fall times. Bi-exponential fitting reveals two components that could be related to free carrier and defect-mediated charge transport. In this scenario, photoexcited carriers passivate defect states, after which the photocurrent is governed by free carriers or shallow traps. For the $n = 2$ phase pellets, the photocurrent stabilizes after approximately 30 s (the overall decrease of the current-time trace is related to a very slow drift of the dark current), while for the $n = 3$ phase pellet the rise in photocurrent continues until the light is switched off. The latter indicates that the photoexcited carriers are not sufficient to passivate all trap states.

The spectral responsivity measured from the pellets is displayed in Figure 4d. For the $n = 1$ phase pellet we obtain a narrow band peak at 420 nm, corresponding to the excitonic band edge of the material. The onset of the photocurrent is red-shifted for the $n = 2$ and $n = 3$ pellets, as expected from the reduced quantum confinement, and the spectral features of the responsivity are in good agreement with the absorption spectra.

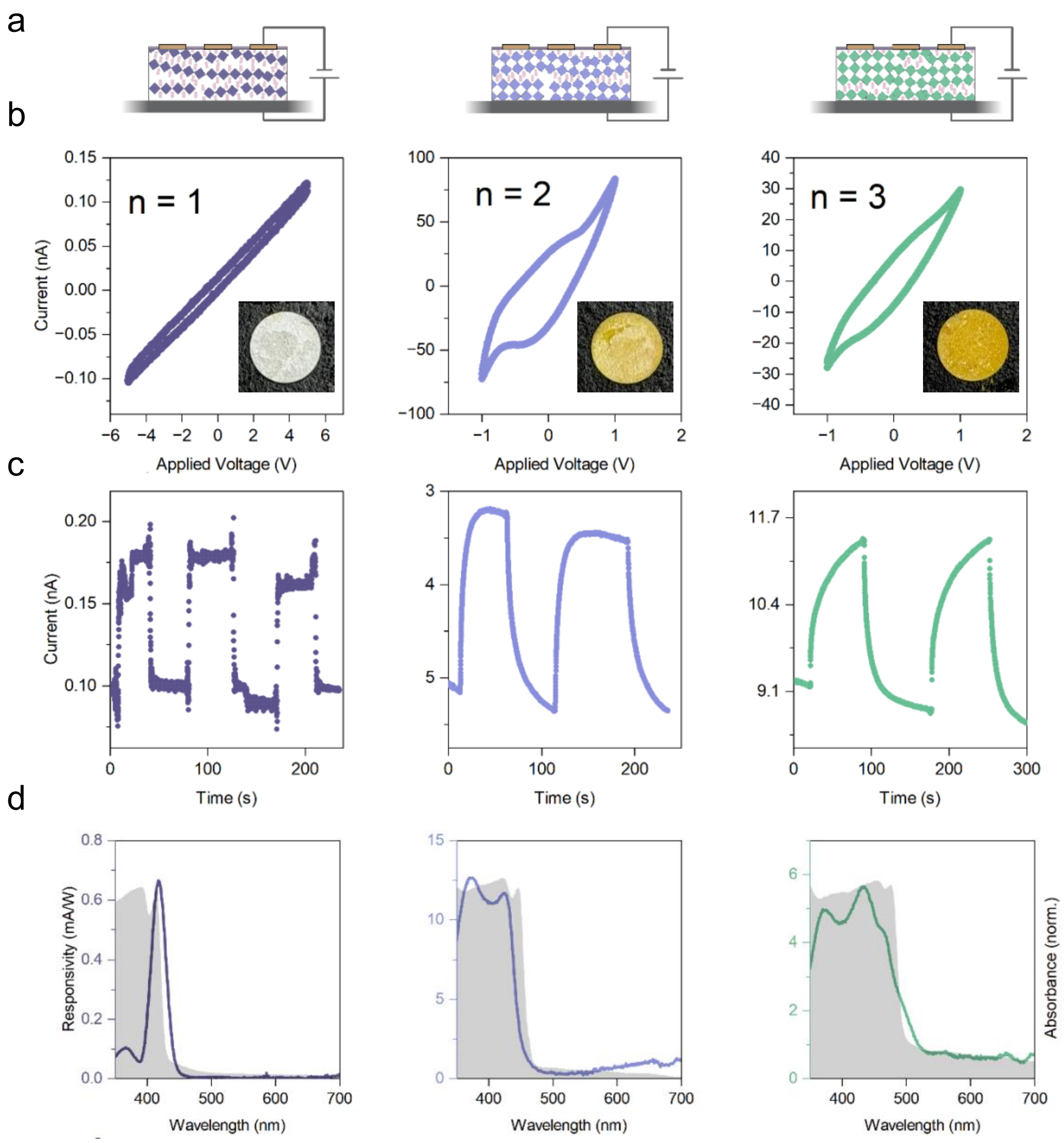

**Figure 4.** Electrical and photoconductive characterization of pellet-based samples fabricated from phase-pure layered perovskites with n = 1–3. a) Schematic illustration of the pellets and the electrical contacts. b) Dark current–voltage (I–V) characteristics, for $n = 1$, $n = 2$, and $n = 3$ pellets. c) Current–time (I–t) measurements under on/off illumination of white light. d) Spectral photo responsivity (colored) overlaid with the corresponding optical absorption spectra (gray). Please see Figure S21 for plots of the spectral detectivity.

To explore the effect of controlled phase mixing on charge transport, we prepared pellets by combining phase-pure materials in a 1:1 ratio. Three representative mixtures were investigated: $n = 1/n = 2$ (Mix1), $n = 1/n = 3$ (Mix2), and $n = 2/n = 3$ (Mix3), as illustrated in Fig. 5a. Elemental analysis of the mixes is presented in Figs. S19 and S20. All three mixed pellets show conductivity and hysteresis in dark current similar to the pure higher $n$ phases (Fig. 5b). However, the time-dependent photoresponse behavior is different (Fig. 5c). Here Mix1 shows a slow photoresponse with a sawtooth pattern in the light on/off current trace that points to the introduction of traps at the microcrystal interfaces. On the other hand, Mix2 and Mix3 feature well-defined photocurrent onsets and decays at the switching events, on a time scale below 1 s (Mix2) or of a few seconds (Mix3). Here the presence of the n=3 phase could lead to more conductive pathways at the edge interfaces of the microcrystals due to the larger inorganic (semiconducting) phase in the cross section. Thus, by mixing, we are able to obtain a faster photoresponse, and the overall behavior indicates that the $n$=1 phase with high organic (dielectric) contribution at the edges is detrimental for the conductive pathways that are dominated by the edges of the microcrystals, while incorporating the $n$=3 phase can improve the conductive response of these channels. As shown in Figure 5d, the spectral responsivity of the mixed pellets reflects an overlay of the absorption features of the different phases, featuring peaks from both phases, and the spectral shape agrees well with the absorbance measured from the mixed pellets. The responsivity spectra reveal a systematic red shift of the photocurrent onset with increasing average layer thickness in the mixtures, consistent with the trend observed for the phase-pure samples. The spectral response closely follows the corresponding optical absorption, indicating that the spectral photoresponse is governed by the intrinsic band-edge transitions of the different phases. This behavior enables the tuning of the color temperature of the pellets that translates also to the spectral photocurrent, and therefore the mixing strategy allows tailoring the spectral absorption and photocurrent profiles. These findings highlight controlled phase mixing as a simple post-synthetic strategy to engineer both the electronic structure and transport behavior of layered perovskites.

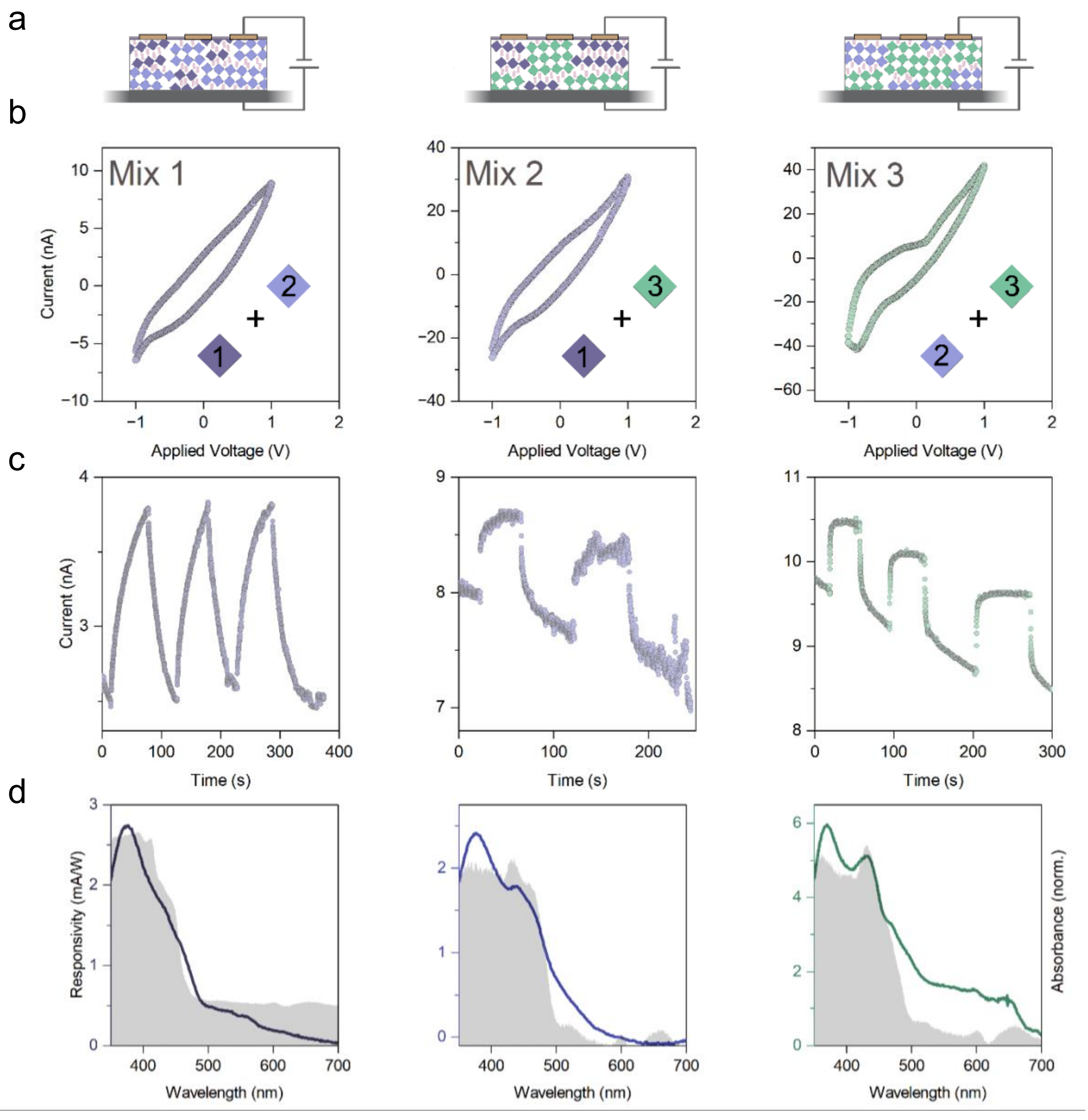

**Figure 5.** Electrical and photoresponse characteristics of pellet-based samples fabricated from controlled mixtures of phase-pure layered perovskites: $n = 1/n = 2$ (Mix1), $n = 1/n = 3$ (Mix2), and $n = 2/n = 3$ (Mix3). a) Schematic representation of mixed-phase pellets highlighting the combinations of different n-phases. b) Current–voltage (I–V) characteristics featuring

reproducible hysteretic behavior for all mixes. c) Current–time (I–t) traces under white light on/off illumination. d) Photoresponsivity spectra (colored) overlaid with the corresponding optical absorption spectra (gray).

## 3. Conclusion

We demonstrated a simple, up-scalable and low-cost synthesis to fabricate two-dimensional metal halide perovskites with controlled and phase-pure *n* number of connected octahedra layers that enable fine control of the optoelectronic properties. After dissolving the precursors at 120 °C the growth process occurs at room temperature over several hours, leading to microcrystal powders with narrow emission linewidths and sharp absorption onset that is defined by the *n*-dependent quantum confinement in the material. We obtain the higher-*n* phases using Cs and BA as A-cations, which extends the range of 2D metal halide perovskite materials in which such confinement control has been achieved. We introduce controlled mixing of powders with different *n* number to design the spectral absorption profile based on the microcrystal powders with phase-pure and stable *n*-phases, and we demonstrate the advantages of such mixing via spectral photocurrent measurements of pellets with mixed composition. We foresee that our approach can be extended to other 2D metal halide perovskites with different composition, and that its simplicity and cost-efficiency can have a strong impact on the application of these materials in optoelectronic applications such as light emitting devices and photodetectors. 2D metal halide perovskites with mixed *n* number are highly advantageous for efficient charge carrier funneling to the lowest band gap phase, and the controlled mixing approach demonstrated here can open novel pathways in this direction.

## 4. Methods

*Chemicals:* Lead(II)bromide ($PbBr_2$, ≥98%), cesium bromide (CsBr, ≥99.9%), hydrobromic acid (HBr, 48%), acetone (≥99.5%), hexane (≥99%) were purchased from Sigma-Aldrich. n-Butylammonium bromide (BABr, ≥98%) was purchased from Great Cell Solar. All chemicals were used without any further purification.

Synthesis of $(BA)_2PbBr_4$ ($n$ = 1) Microcrystalline Powder:

$(BA)_2PbBr_4$ microcrystalline powder was synthesized using a previously published approach.[41] Specifically, $PbBr_2$ (0.25 mmol, 92 mg) was dissolved in 200 µL of HBr under vigorous stirring and then diluted with 2 mL of acetone. Subsequently, BA (0.6 mmol, 59 µL) was injected into the solution, leading to the immediate formation of micrometer-sized $(BA)_2PbBr_4$ microcrystals. The reaction mixture was stirred overnight to ensure complete reaction. Afterward, the microcrystalline powder was isolated from the solution by centrifugation (3000 rpm, 3 min) and redispersed in 2 mL of acetone. The washing procedure was repeated two additional times. Finally, the purified powder was dried overnight in a lyophilizer after preliminary freezing at −80 °C.

Synthesis of $(BA)_2CsPb_2Br_7$ and $(BA)_2Cs_2Pb_3Br_{10}$ ($n$ = 2, 3) Microcrystalline Powders:

$(BA)_2CsPb_2Br_7$ and $(BA)_2Cs_2Pb_3Br_{10}$ microcrystalline powders were synthesized using precursor ratios from a previously published protocol;[39] however, the procedure itself was significantly modified to obtain phase-pure products and achieve reproducible results.

For $n$ = 2, the optimized molar ratio of BA/Cs/Pb was 3:1:2, while for $n$ = 3 it was 3.5:2:3. The excess BA relative to the ideal stoichiometry was necessary to stabilize the layered phases and suppress the formation of bulk perovskite impurities. All precursors were dissolved in the minimum volume of HBr to achieve saturation as rapidly as possible upon cooling, thereby ensuring reproducible crystallization and reaction conditions.

Specifically, for n = 2 phase, BABr (0.75 mmol, 115.5 mg), CsBr (0.25 mmol, 53.2 mg), and $PbBr_2$ (0.5 mmol, 183.5 mg) were dissolved in 2.5 mL of saturated aqueous HBr. For n = 3 phase, BABr (0.875 mmol, 134.8 mg), CsBr (0.5 mmol, 106.4 mg), and $PbBr_2$ (0.75 mmol, 275.3 mg) were dissolved in 3.5 mL of saturated aqueous HBr. The volume of HBr was determined experimentally for each *n* value to ensure complete dissolution of the precursors while maintaining a highly concentrated solution. More diluted solutions did not crystallize efficiently and resulted in poor phase selectivity.

The formation of phase-pure single crystals was achieved by optimizing the precursor stoichiometry, reaction temperature, and cooling conditions. Under the optimized protocol, the precursors were completely dissolved by heating the reaction mixture at 120 °C under continuous stirring for approximately 30 min. The solution was then placed in a water bath at 20 °C for 17 h, resulting in the formation of yellowish microcrystals. The crystals were collected by suction filtration using hexane as a nonpolar solvent to wash and transfer the crystals while preserving phase purity. The obtained powder was subsequently dried overnight in a lyophilizer after preliminary freezing at −80 °C for further use. However, it should be noted that this approach requires careful handling of the final product, as any disturbance of the solution equilibrium may immediately lead to irreversible phase transformation and the formation of bulk perovskite phases in the sample, which complicates the isolation of a phase-pure product.

In-Situ PL Measurements:

*In-situ* PL measurements were performed during the crystallization process at temperatures ranging from 3 to 60 °C. The precursor solution was placed in a 3 mL quartz cuvette immersed in a temperature-controlled water bath. Excitation was provided by a 365 nm LED (Thorlabs) coupled to a bifurcated optical fiber. The fiber tip was positioned immediately next to the cuvette wall and immersed in the water bath to minimize thermal gradients and maintain stable measurement conditions.

The emitted PL was collected through the same fiber and detected using a fiber-coupled spectrometer (AvaSpec-2048, Avantes). Prior to detection, the collected signal was passed through a 400 nm long-pass filter to suppress excitation light. The spectrometer integration time was set to 200 ms.

The excitation beam was slightly defocused to probe a larger volume and minimize intensity fluctuations caused by light scattering from the forming microcrystals. The flask was continuously stirred to prevent sedimentation and accumulation of microcrystals at the bottom and around the probe.

Optical Characterization:

Steady-state photoluminescence (PL) spectra were acquired using an Edinburgh Instruments (FLS920) fluorescence spectrometer equipped with a xenon lamp and monochromator. Time-resolved PL measurements were carried out using a time-correlated single-photon counting (TCSPC) system coupled to a pulsed diode laser (371 nm, 50 ps pulse width, 1 MHz repetition rate).

Absorption spectra were collected using a Varian Cary 5000 UV–Vis–NIR spectrophotometer equipped with an external diffuse reflectance accessory, operated in absorption mode. Hyperspectral confocal imaging was performed using a Nikon A1R+/A1+ confocal laser microscope equipped with a 20× objective. The samples were excited at 400 nm, and PL was collected in the 410–530 nm range with a spectral resolution of 2.5 nm.

Device fabrication and characterization:

Pellets were prepared using a 13 mm pellet die set (Specac, UK) under a pressure of 200 MPa. Approximately 50 mg of perovskite powder was loaded into the die to form a pellet. To reduce the pellet thickness, the powder was pressed directly onto the substrate, allowing the pellet to be transferred to a KE500ET combined e-beam/thermal evaporator (Kenosistec, Italy) for deposition of 50 nm thick Au electrodes. A two-electrode configuration was employed for all measurements. The electrodes were deposited through a shadow mask.

Current–voltage (I–V) and current–time (I–t) characteristics were measured using a Keithley 2612A sourcemeter in combination with a micromanipulator probe station (Janis Research) and a fiber coupled white light source (Newport 66902 Xe lamp) that for the spectral measurements was couple to a monochromator (Spectral Products CM110). The lamp spectrum was measured with a DET-10A photodiode from Thorlabs. All measurements were carried out in ambient air at atmospheric pressure

Acknowledgements

This work has received funding by the European Union under Project 101131111 – DELIGHT.

Received: ((will be filled in by the editorial staff))

Revised: ((will be filled in by the editorial staff))

Published online: ((will be filled in by the editorial staff))

**Supporting Information**

Supporting Information is available from the Wiley Online Library or from the author.

Table of contents:

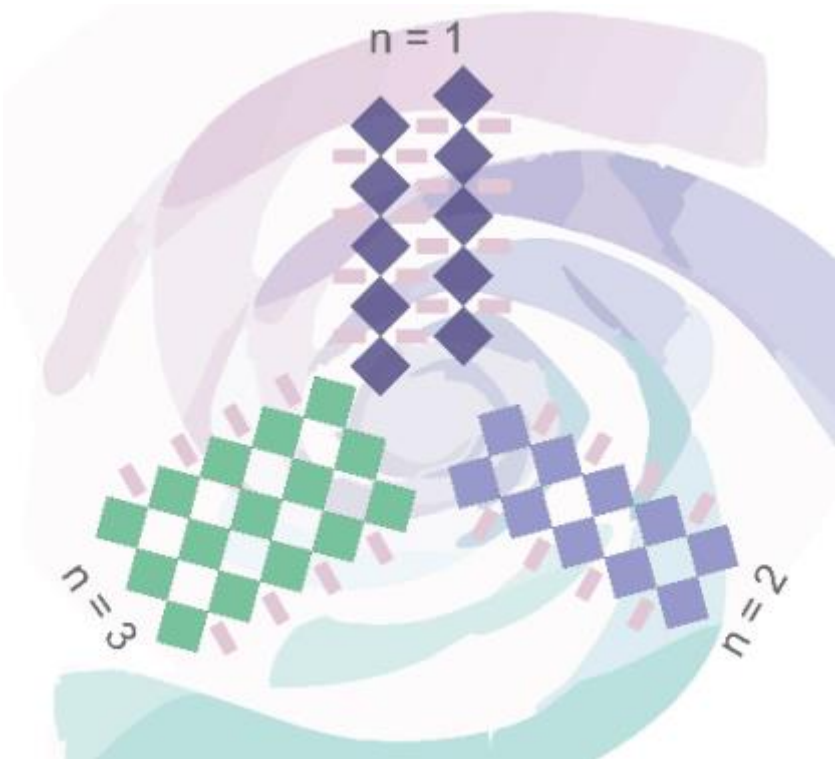

Phase purity in layered lead bromide perovskites is achieved by controlling the number of inorganic layers (n) through a scalable synthesis approach. Tunable optical absorption is obtained by mixing phase-pure powders, enabling precise spectral adjustment. The resulting materials exhibit composition-dependent optical responses, highlighting their potential for applications in band-edge engineering.

# Supporting Information

## Confinement Controlled Phase-Pure 2D Layered Perovskites for Band-Edge Engineering

*Danila A. Tatarinov[1,2], Alexander Schleusener[1], and Roman Krahne[1]*

[1]*Optoelectronics, Italian Institute of Technology (IIT), Via Morego 30, 16163 Genoa, Italy*
[2]*Dipartimento di Chimica e Chimica Industriale, Università degli Studi di Genova, Via Dodecaneso, 31, Genova 16146, Italy*

*Corresponding authors email: roman.krahne@iit.it*

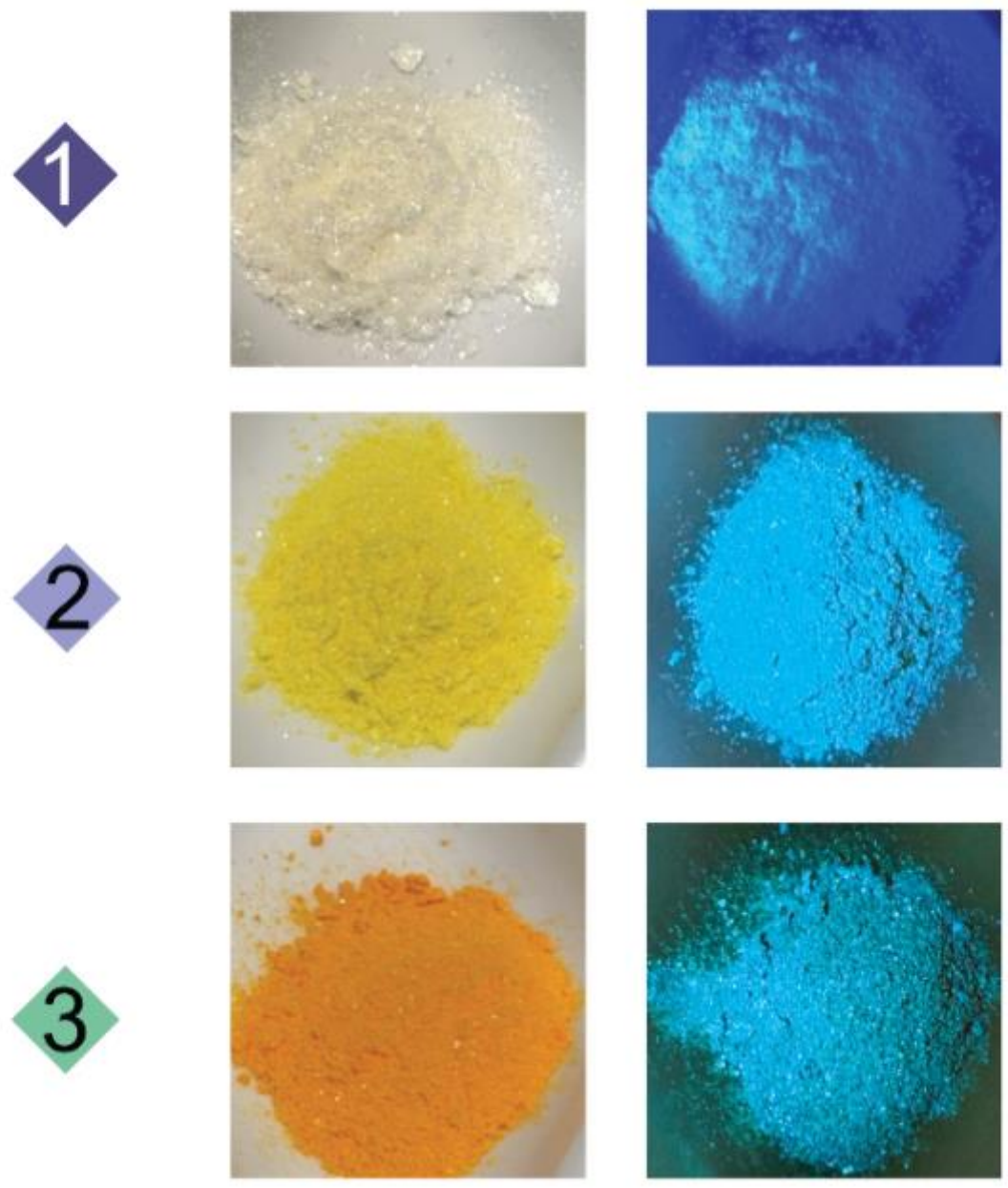

**Figure S1.** Photographs of the phase-pure layered perovskite microcrystals ($n$ = 1–3) under ambient illumination (left) and UV excitation (365 nm, right).

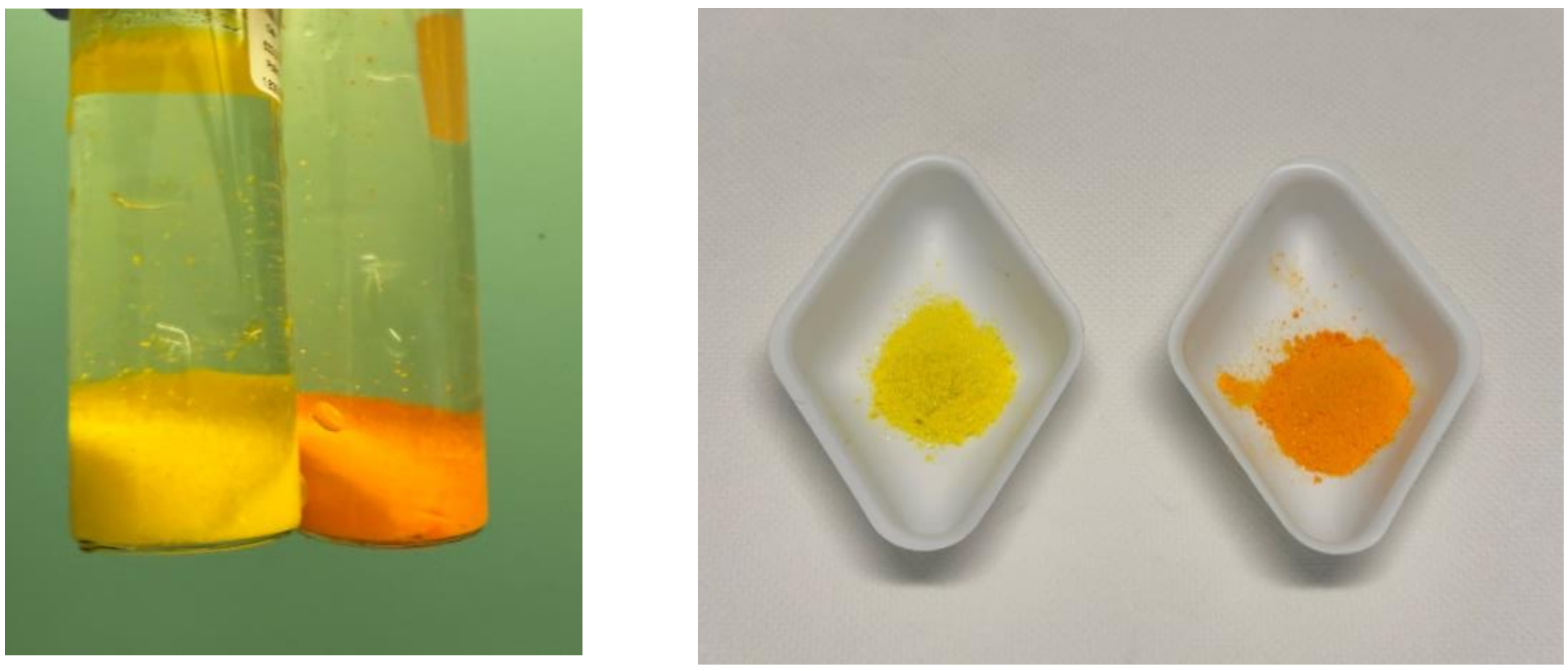

**Figure S2.** Gram-scale synthesis of phase-pure layered perovskites $n = 2$ and $n = 3$. Photographs of the products obtained from a 10-fold scale-up synthesis (1.5 g per batch) before isolation (left) and as dried powders (right), demonstrating the scalability of the synthetic protocol.

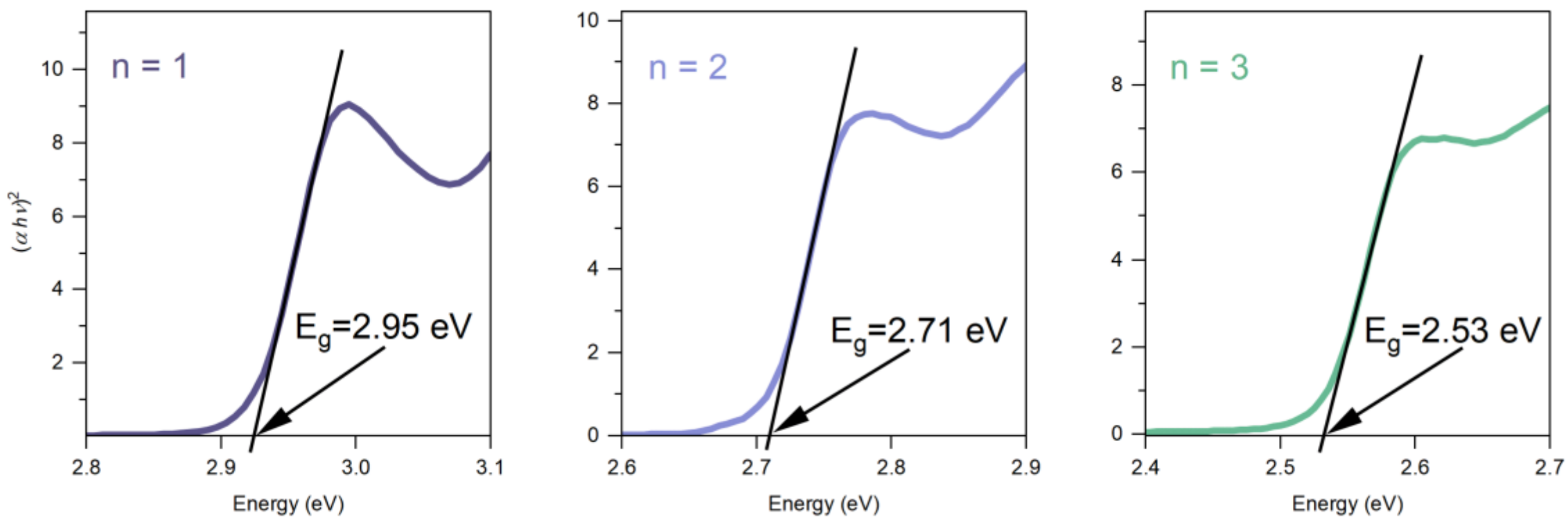

**Figure S3.** Tauc plots for 2D perovskite samples with *n* = 1, 2, 3. The optical bandgap $E_g$ values are determined by linear extrapolation of the absorption edge.

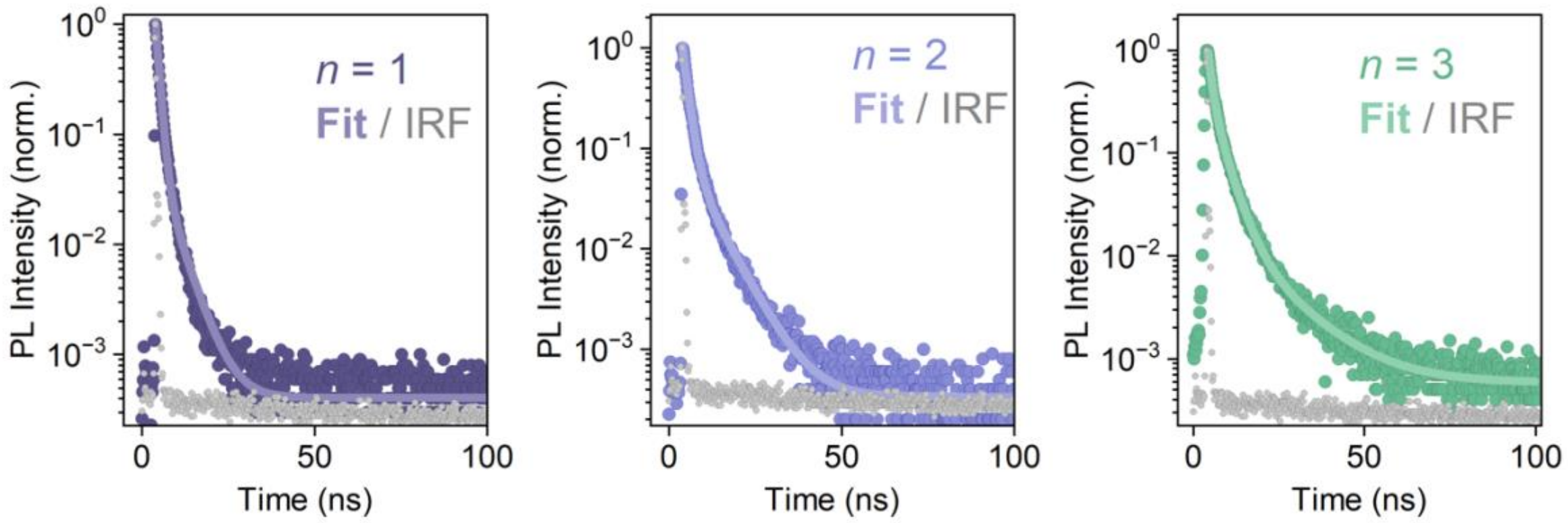

**Figure S4.** TRPL decay curves of phase-pure 2D layered perovskites (*n* = 1–3). Symbols represent experimental data, solid lines the reconvolution fits, and gray symbols the instrument response function (IRF).

**Table S1.** Photoluminescence decay parameters obtained from biexponential fitting of 2D perovskite samples with *n* = 1, 2, 3.

| Sample | Emission (nm) | $A_1$ | $\tau_1$ (ns) | $A_2$ | $\tau_2$ (ns) | $A_3$ | $\tau_3$ (ns) | $\tau_{iwa}$ (ns) |
|---|---|---|---|---|---|---|---|---|
| *n* = 1 | 410 | 0.957 | 1.006 | 0.043 | 4.556 | - | - | 4.90 |
| *n* = 2 | 442 | 0.9945 | 5.78 | 0.0021 | 46.32 | 0.0033 | 167.44 | 6.40 |
| *n* = 3 | 473 | 0.1950 | 0.06 | 0.3689 | 2.45 | 0.4362 | 50.36 | 22.89 |

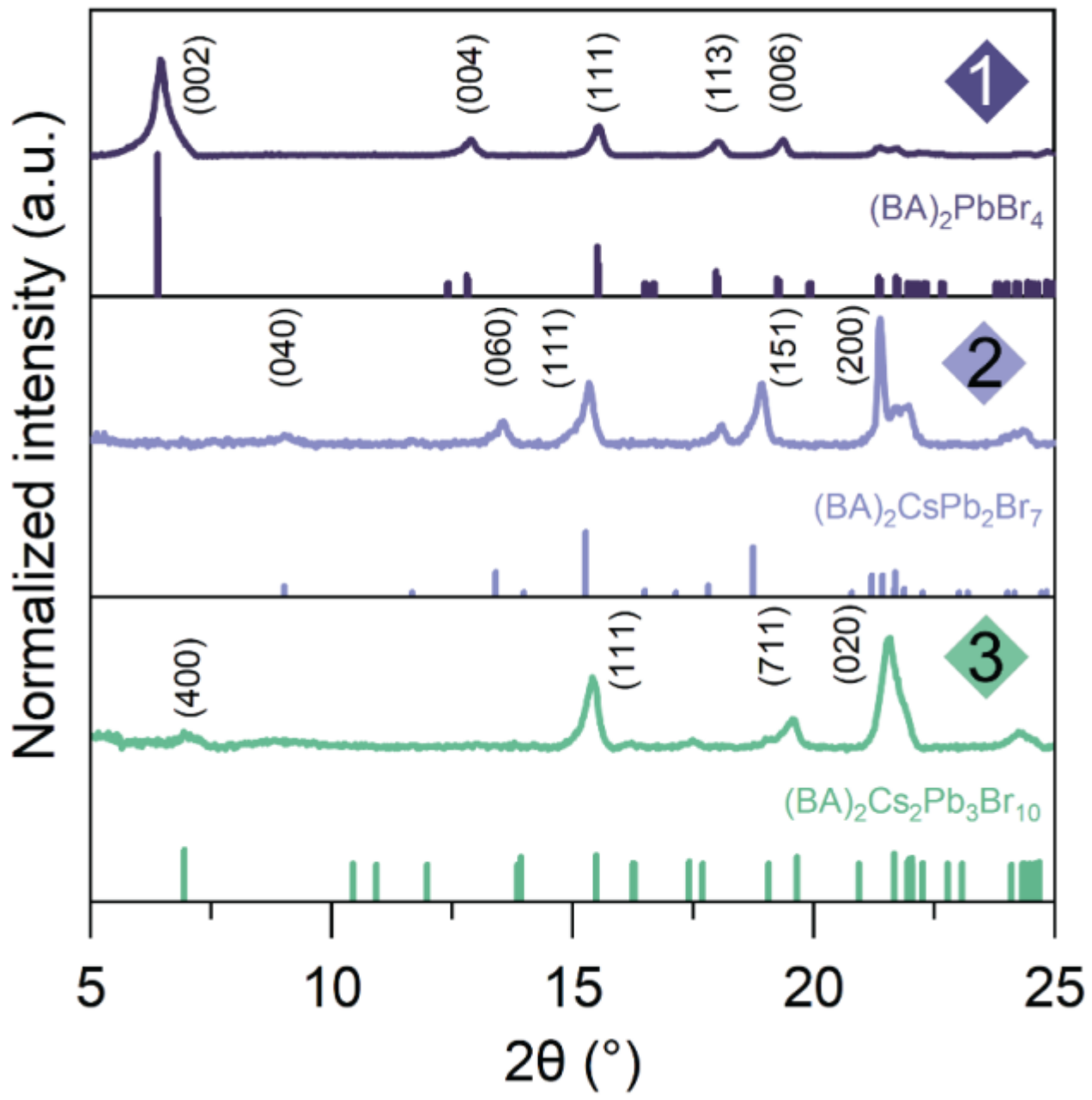

**Figure S5.** Experimental XRD patterns of the phase-pure layered perovskites together with the corresponding simulated diffraction patterns.

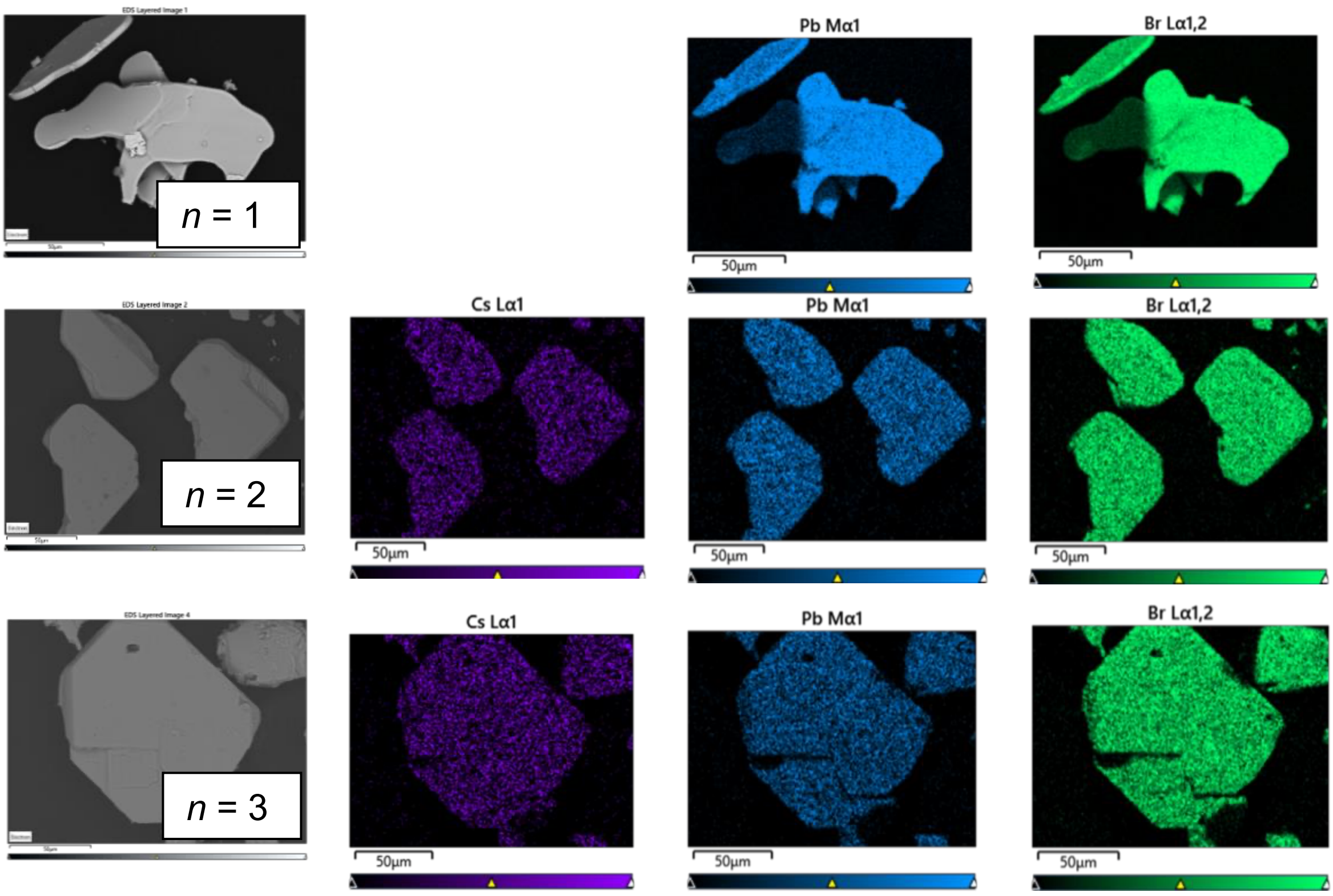

**Figure S6.** SEM images and corresponding EDS elemental maps of representative 2D perovskite microcrystals ($n$ = 1–3), showing the spatial distribution of Cs, Pb, and Br.

n = 1

| Element | Wt% | Atomic % |
|---|---|---|
| Pb | 41.14 | 21.23 |
| Br | 58.86 | 78.77 |
| Total | 100.00 | 100.00 |

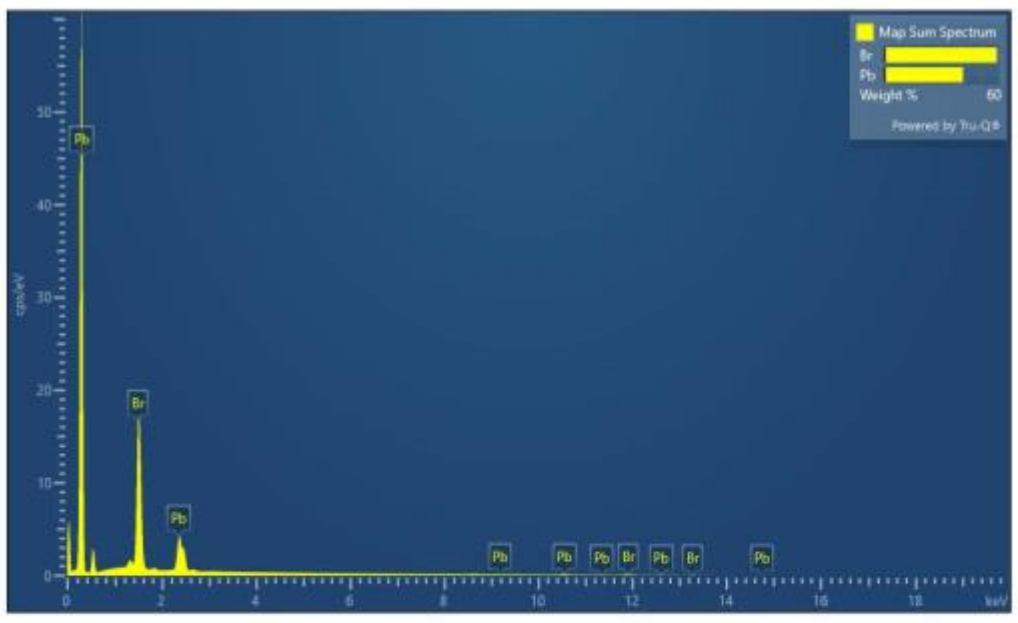

n = 2

| Element | Wt% | Atomic % |
|---|---|---|
| Pb | 38.07 | 20.45 |
| Cs | 12.05 | 10.09 |
| Br | 49.88 | 69.46 |
| Total | 100.00 | 100.00 |

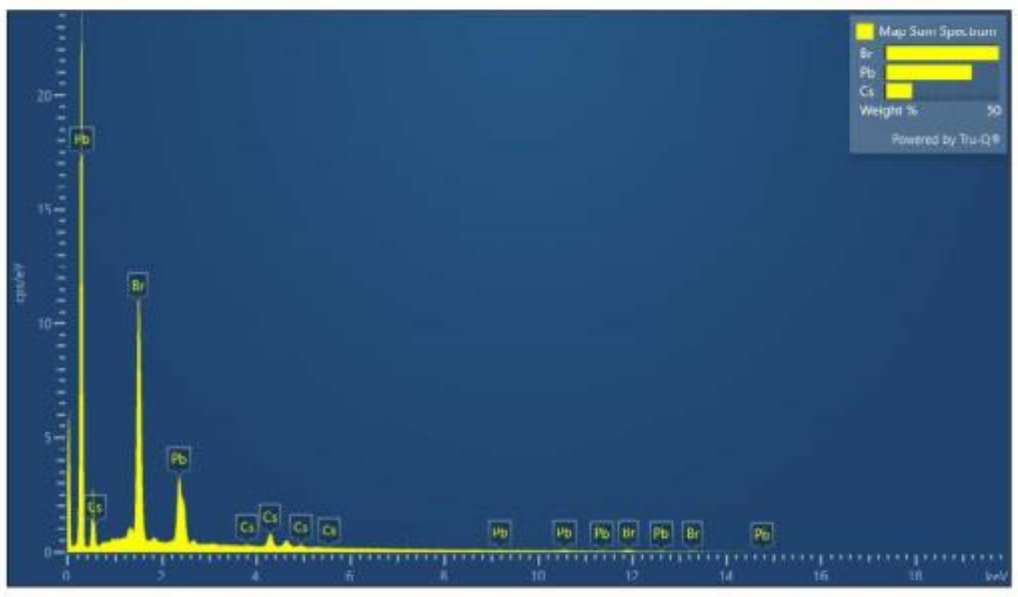

n = 3

| Element | Wt% | Atomic % |
|---|---|---|
| Br | 44.38 | 63.29 |
| Cs | 19.87 | 17.04 |
| Pb | 35.76 | 19.67 |
| Total | 100.00 | 100.00 |

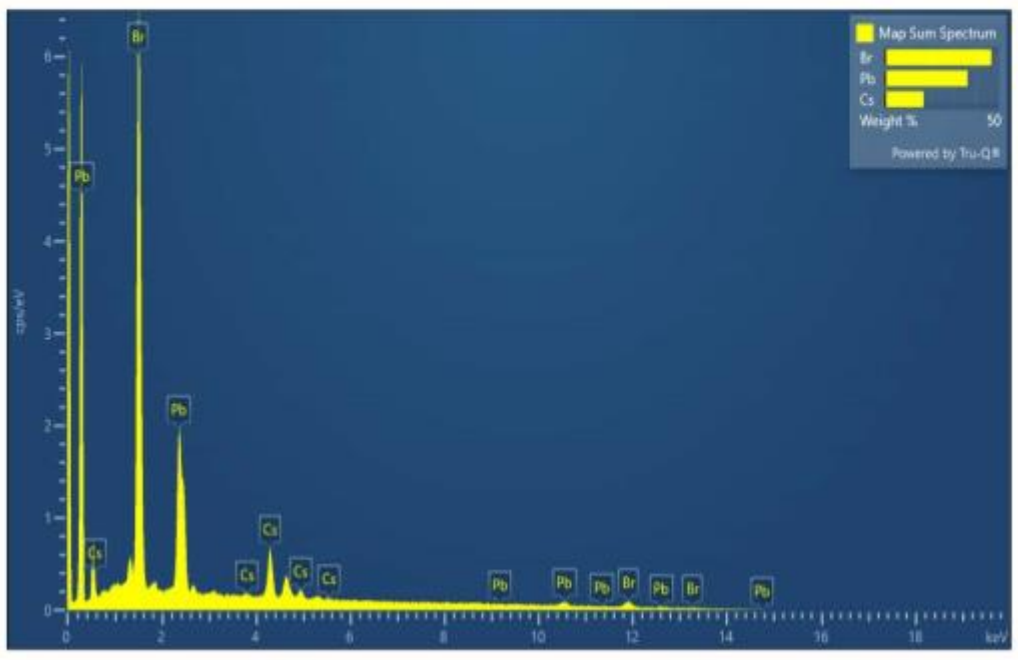

**Figure S7.** EDS spectra and corresponding elemental composition of 2D perovskite microcrystals ($n$ = 1–3), confirming the presence of Pb, Cs, and Br.

**Fitting of the temperature decay of the reaction solution**

Solution temperature decay was analyzed using a biexponential model to account for the presence of fast and slow thermal processes. The temporal evolution of temperature was fitted with the following function:

$$T(t) = T_\infty + A_1 \exp(-t/\tau_1) + A_2 \exp(-t/\tau_2),$$

where $T_\infty$ is the final equilibrium temperature, $A_1$ and $A_2$ are the amplitudes of the fast and slow components, and $\tau_1$ and $\tau_2$ are the corresponding characteristic relaxation times.

Experimental temperature–time data were fitted using nonlinear least-squares minimization. Initial parameters were estimated from the experimental curves and refined iteratively until convergence.

The biexponential model was chosen as it provided an excellent description of the data ($R^2 >$ 0.99 for all samples) and reflects the coexistence of multiple thermal relaxation pathways, which can be attributed to different heat transfer mechanisms such as rapid surface cooling and slower bulk thermal equilibration.

The extracted fitting parameters together with the calculated initial cooling rates are summarized in **Table S2:**

| **Type** | **$T_\infty$ (°C)** | **$A_1$** | **$\tau_1$ (s)** | **$A_2$** | **$\tau_2$ (s)** | **$R^2$** |
|---|---|---|---|---|---|---|
| Water, 3 °C | 1.91 | 75.54 | 22.31 | 62.05 | 64.31 | 0.991 |
| Water, 15 °C | 15.34 | 119.49 | 41.82 | 62.19 | 41.88 | 0.996 |
| Water, 20 °C | 19.83 | 86.39 | 25.42 | 30.83 | 105.40 | 0.994 |
| Water, 35 °C | 35.09 | 113.48 | 19.45 | 19.19 | 72.64 | 0.998 |
| Water, 45 °C | 44.52 | 88.29 | 28.85 | 2.67 | 28.86 | 0.994 |
| Water, 60 °C | 56.52 | 28.26 | 32.72 | 72.40 | 32.72 | 0.997 |
| Cold room, 5 °C | 3.09 | 5.88 | 45.64 | 117.28 | 340.52 | 0.999 |

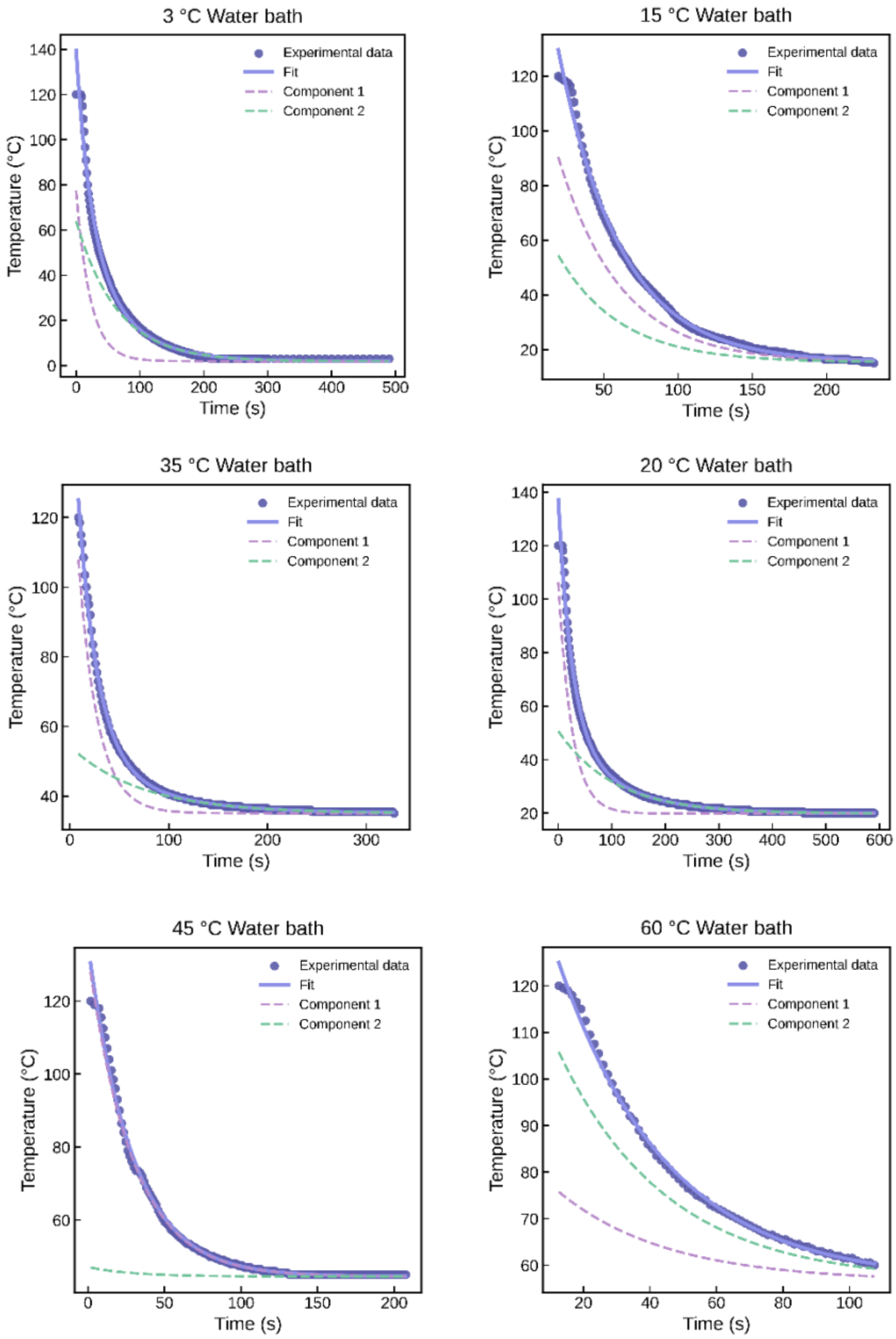

**Figure S8.** Cooling dynamics of the reaction system under different conditions. Experimental temperature profiles are fitted with a biexponential decay model, highlighting fast and slow cooling components under each condition.

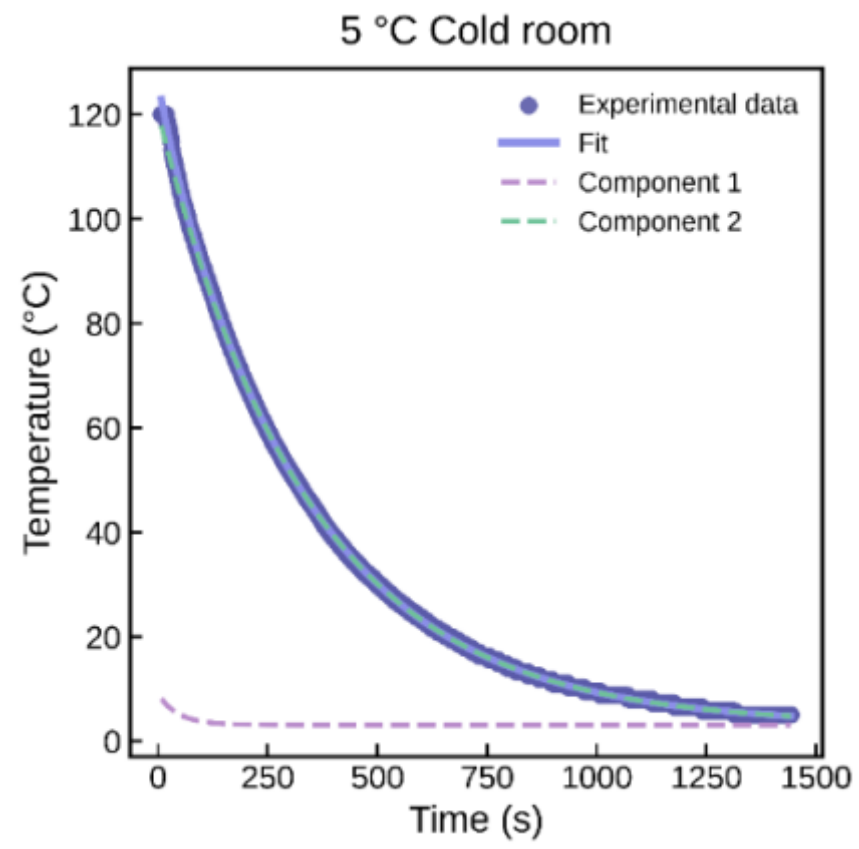

**Figure S9.** Cooling dynamics of the reaction system under cooling in a 5 °C cold room. This type of cooling required the longest time to reach thermal equilibrium among all investigated cooling conditions.

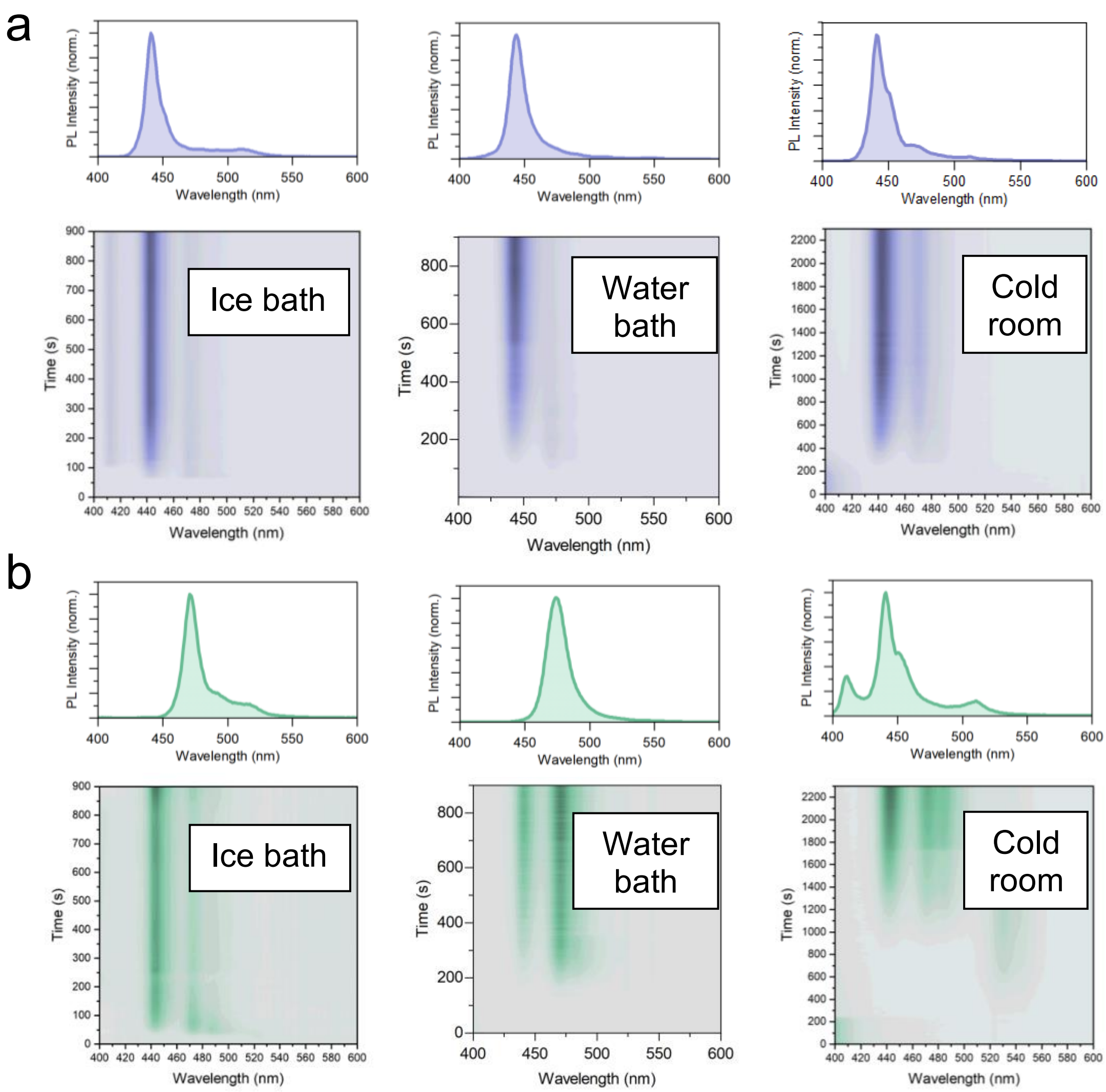

**Figure S10.** PL evolution maps for $n$ = 2 (a) and $n$ = 3 (b) samples under different cooling conditions: ice bath, water bath, and ambient (cold room). The upper panels display the corresponding PL spectra after 12 h of reaction at constant temperature.

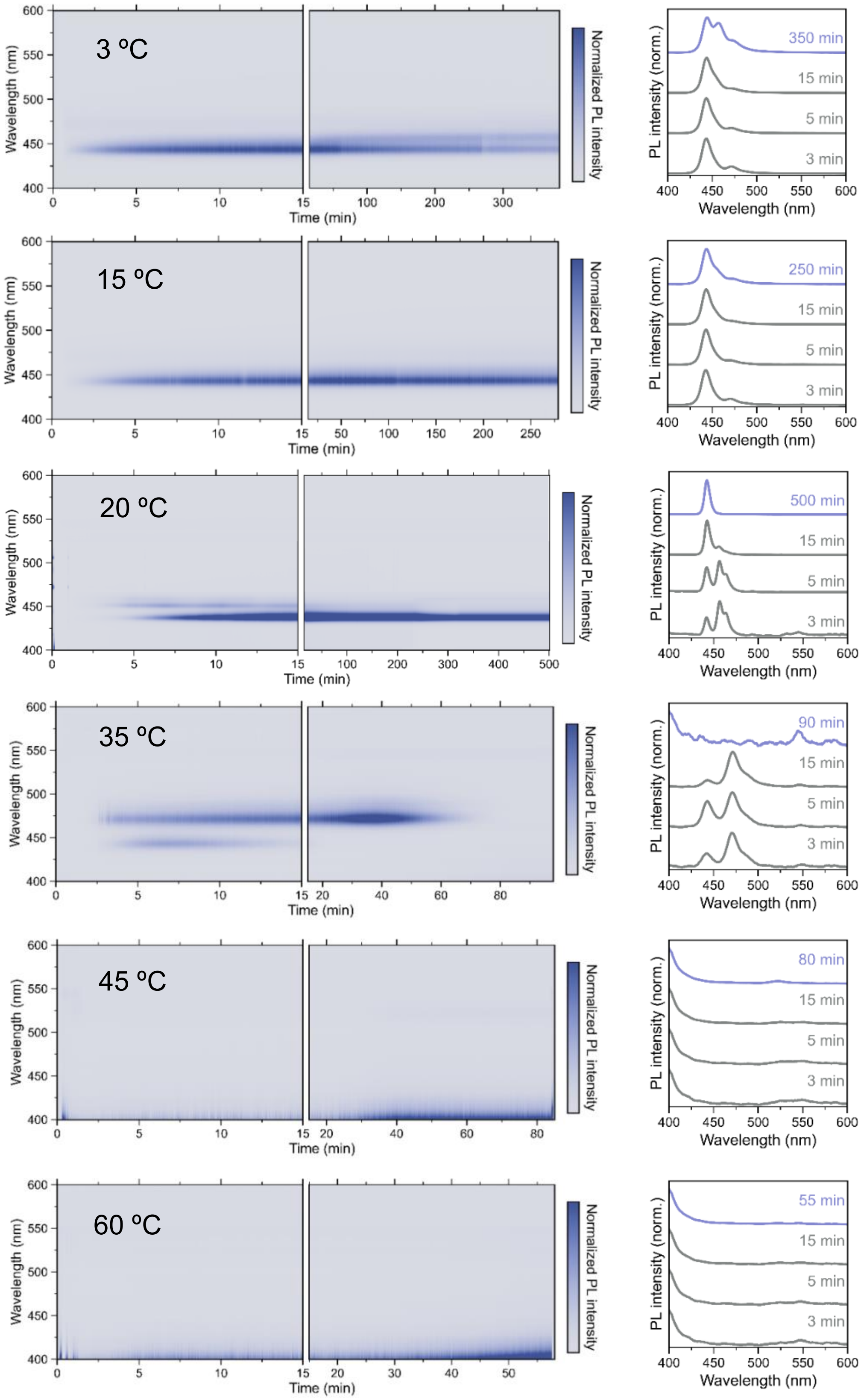

**Figure S11.** Time-resolved PL evolution of the 2D perovskite samples ($n$ = 2) at different crystallization temperatures. Normalized PL intensity contour maps (left) illustrate the spectral evolution during crystallization, while representative normalized PL spectra (right) recorded at selected time intervals highlight the evolution of the spectral emission features.

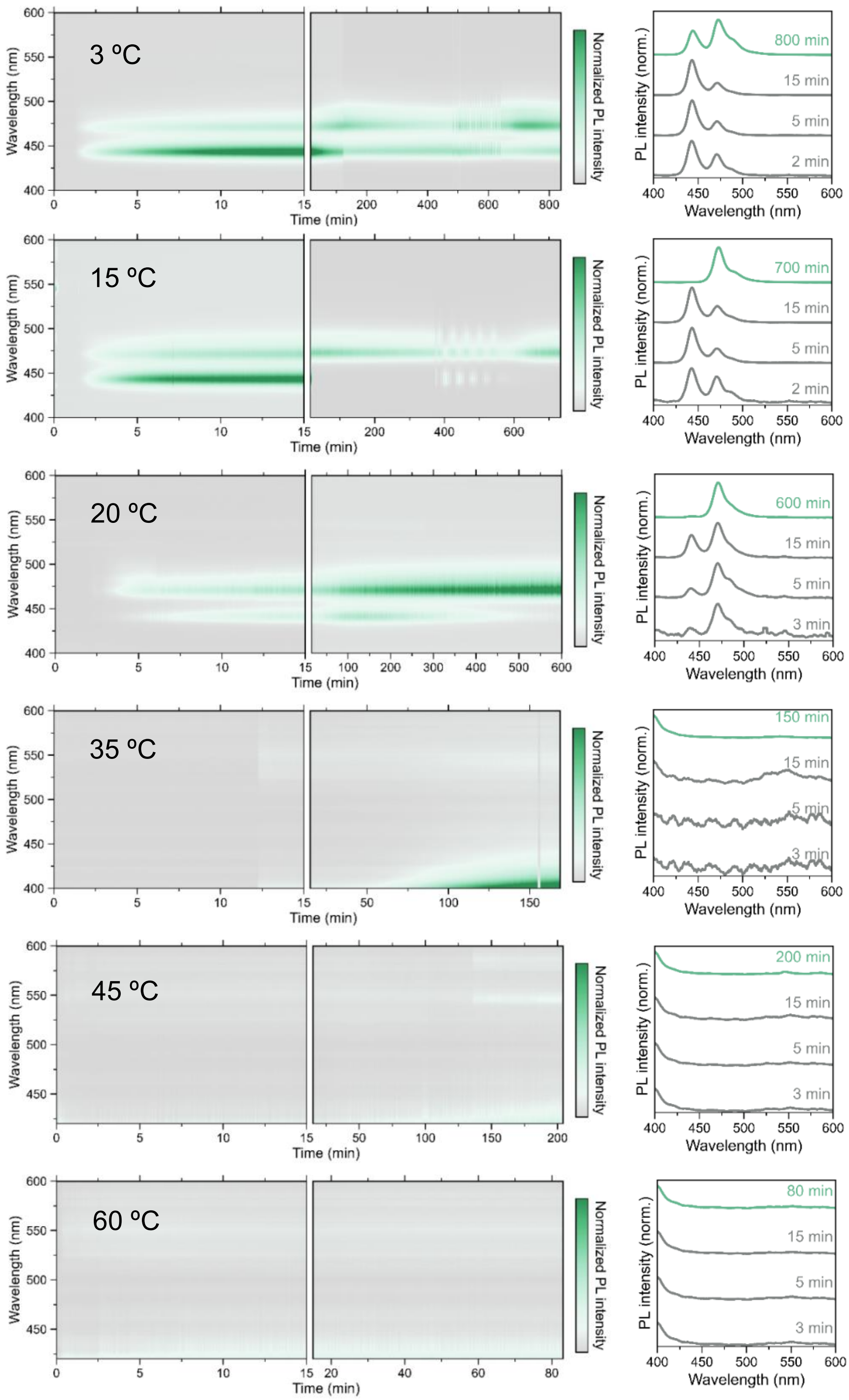

**Figure S12.** Time-resolved PL evolution of the 2D perovskite samples ($n$ = 3) at different crystallization temperatures. Normalized PL intensity contour maps (left) illustrate the spectral evolution during crystallization, while representative normalized PL spectra (right) recorded at selected time intervals highlight the evolution of the spectral emission features.

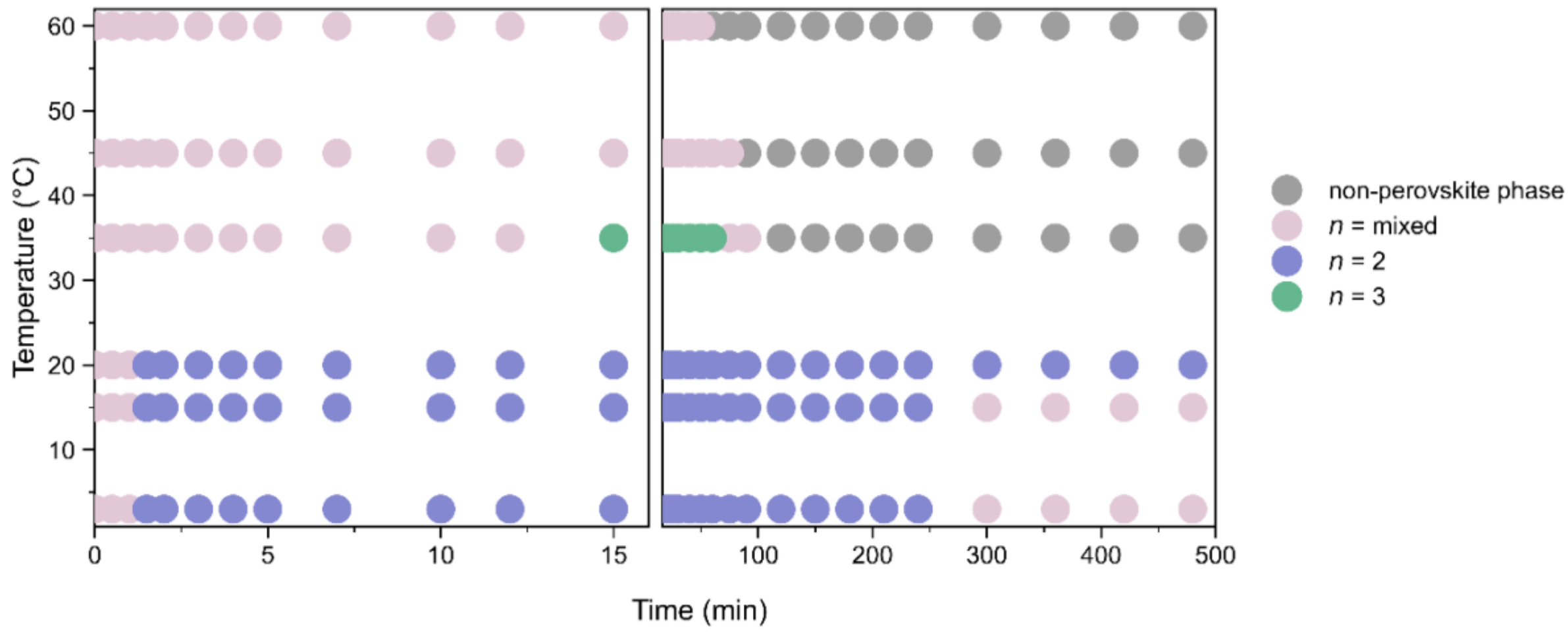

**Figure S13.** Phase classification map derived from in situ PL spectroscopy for the synthesis of the $n$ = 2 2DLPs as a function of synthesis temperature and reaction time. Colors indicate the formation of the predominant phase (>50 %), mixed phases, or non-perovskite products.

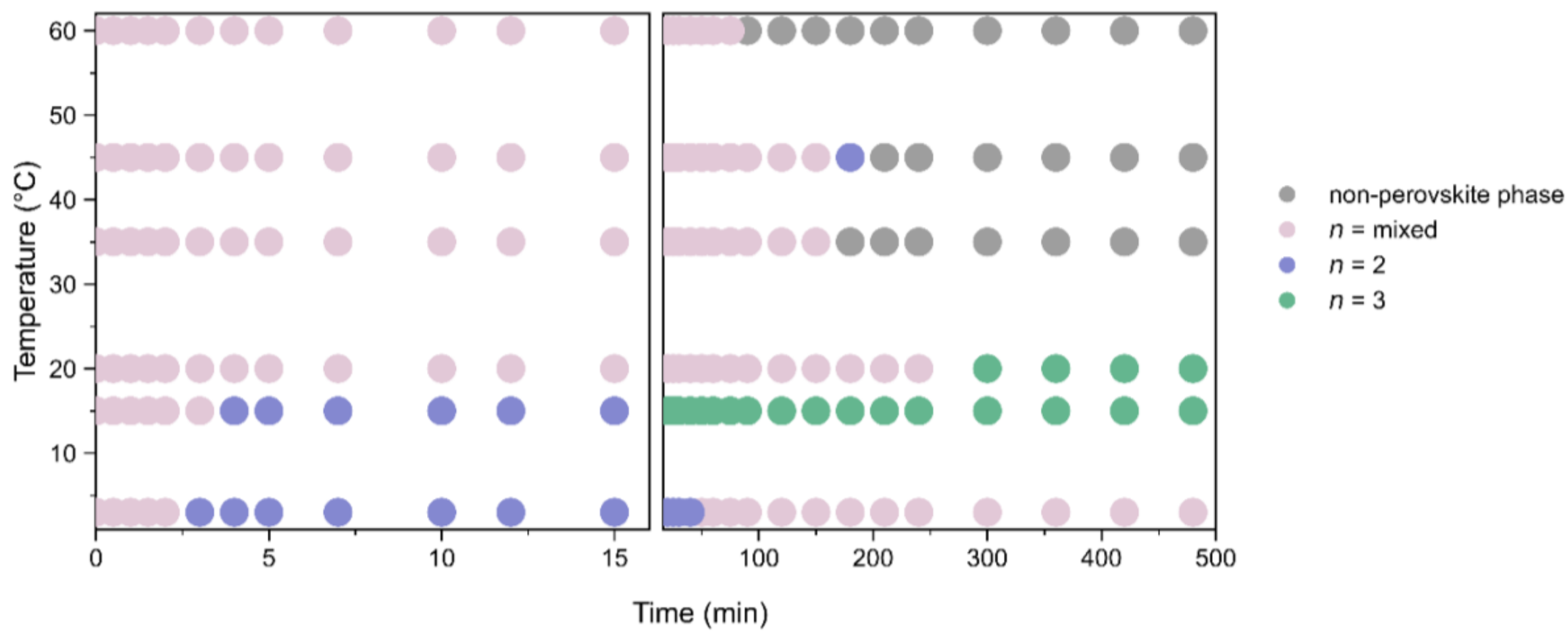

**Figure S14.** Phase classification map derived from in situ PL spectroscopy for the synthesis of the $n$ = 3 2DLPs as a function of synthesis temperature and reaction time. Colors indicate the formation of the predominant phase (>50 %), mixed phases, or non-perovskite products.

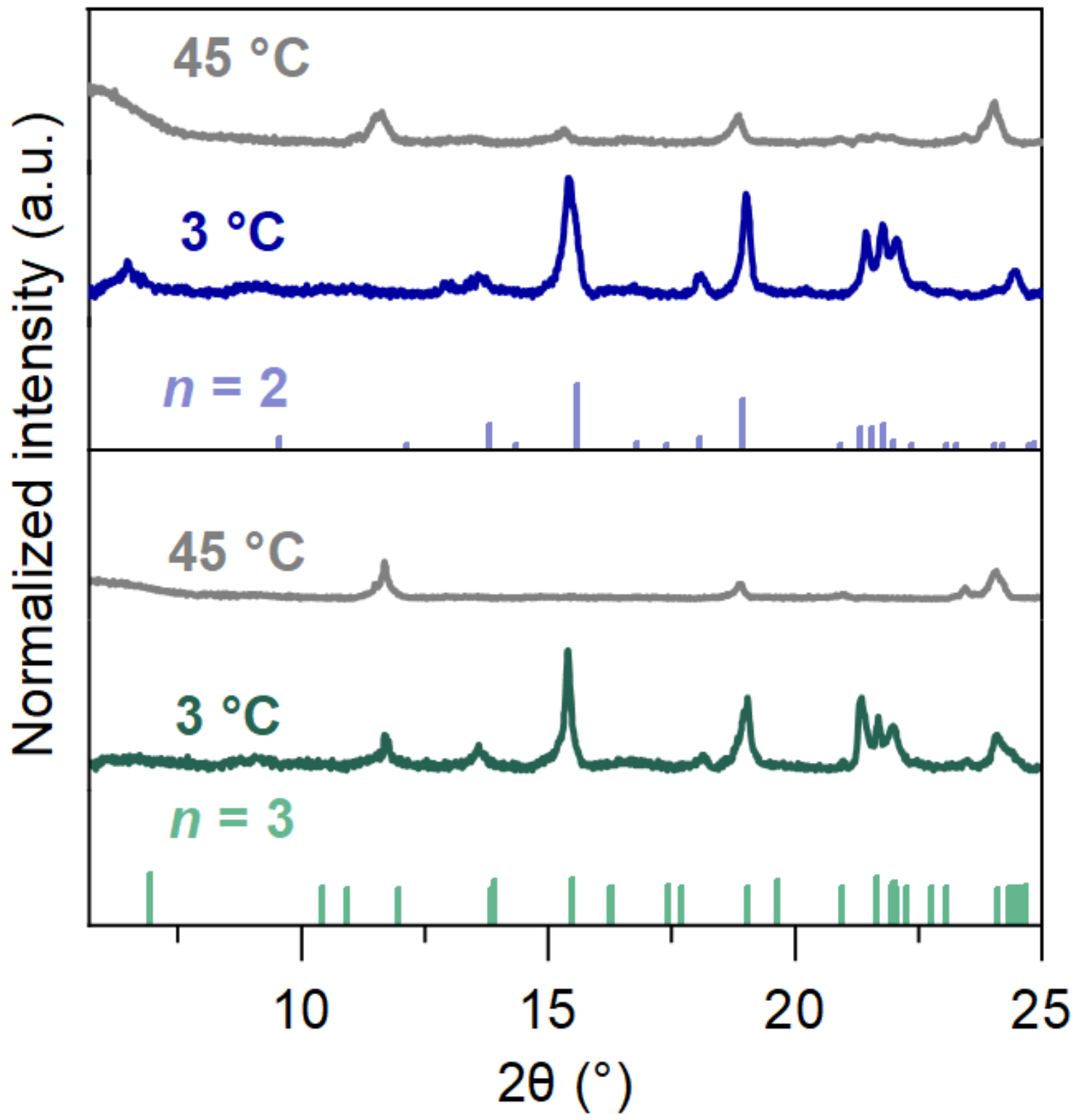

**Figure S15.** Experimental XRD patterns of the products obtained under different synthesis temperatures (3 and 45 °C) for the targeted $n$ = 2 and $n$ = 3 samples. Simulated diffraction patterns of the target phases are shown for comparison.

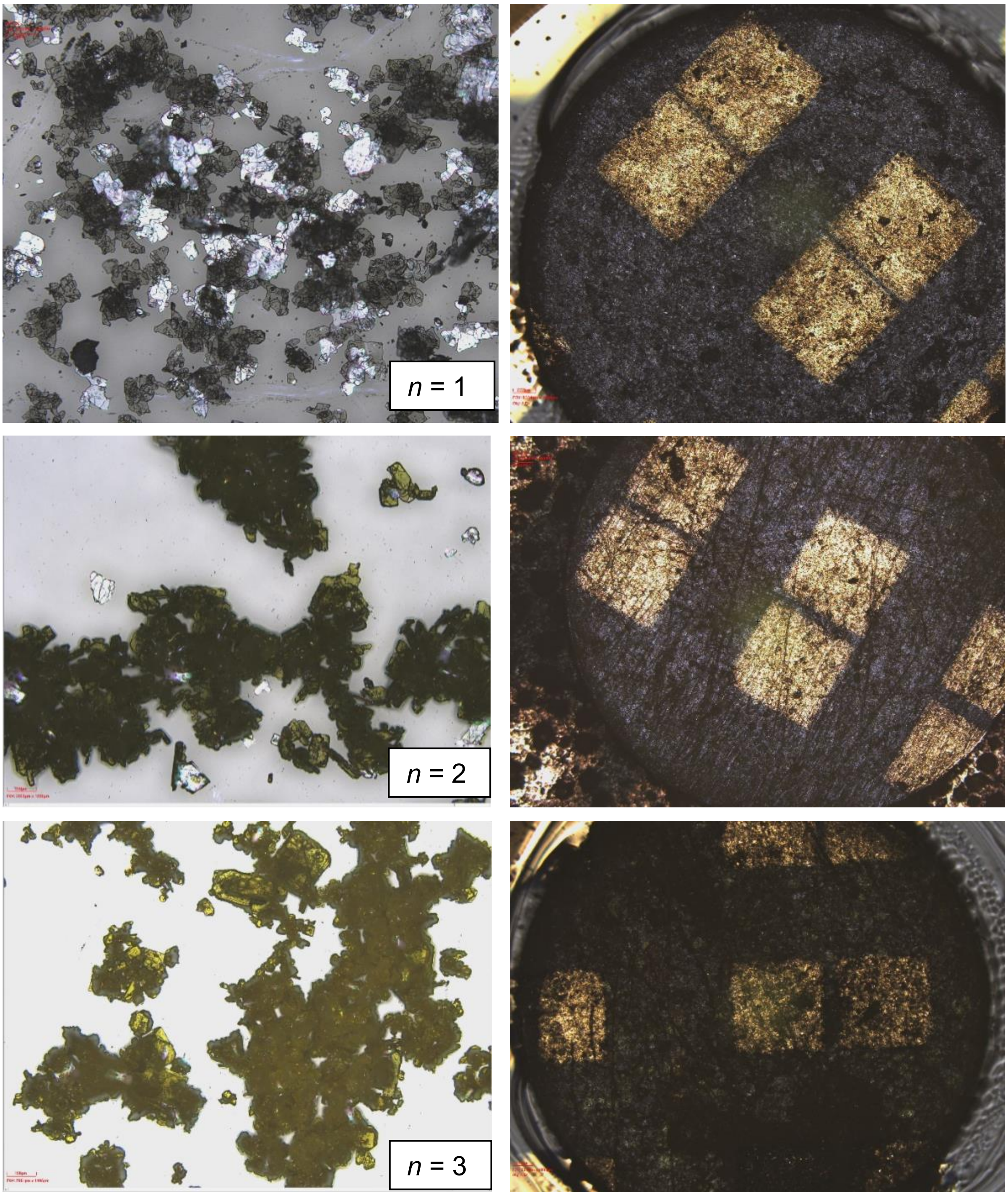

**Figure S16.** Optical microscopy images of 2D perovskite microcrystal powders (left) and corresponding pellet-based devices with Au electrodes (right) for samples with $n$ = 1–3.

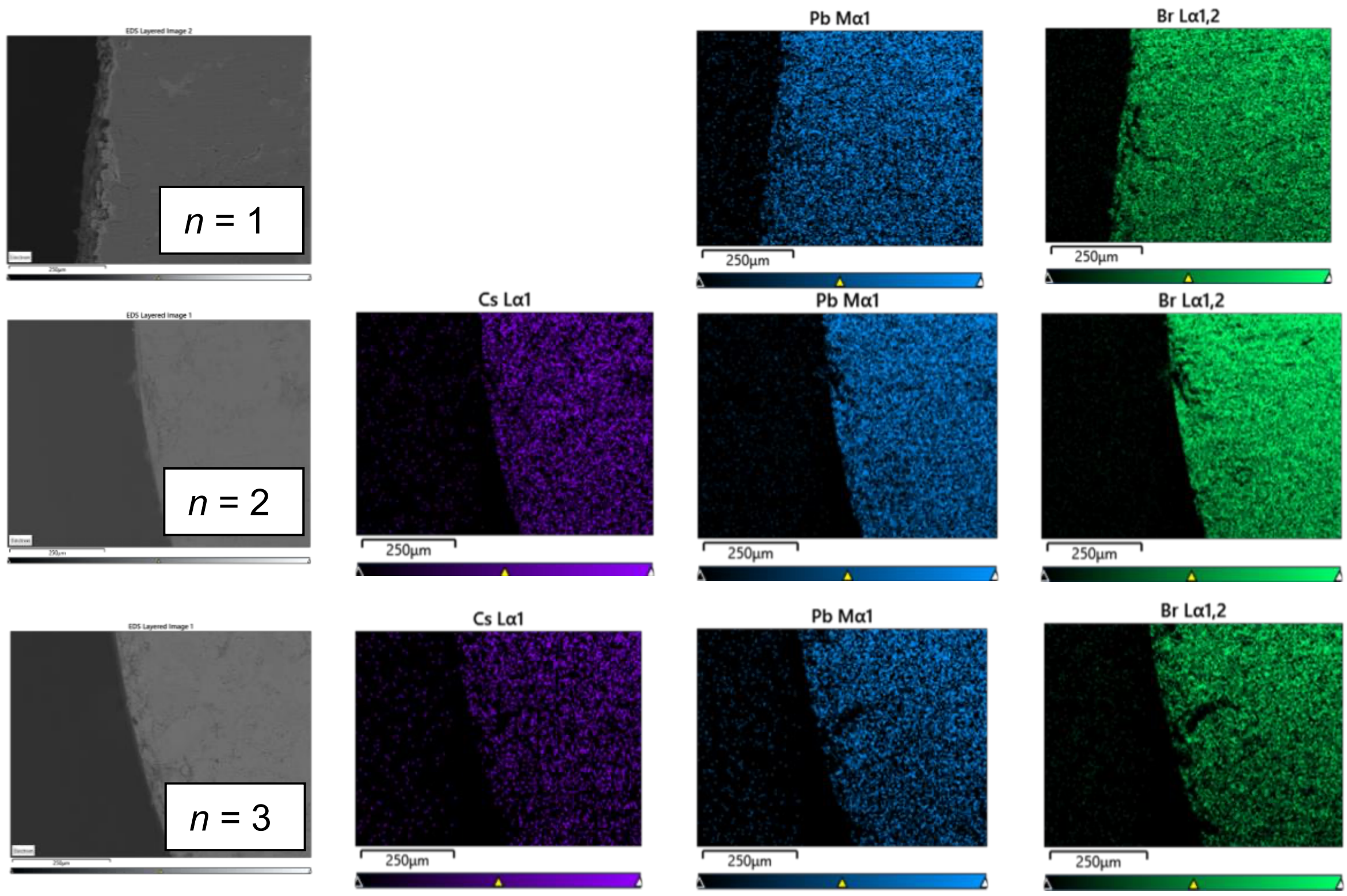

**Figure S17.** SEM images and corresponding EDS elemental maps of the pellets pressed from the pure 2D perovskite microcrystals samples (*n* = 1–3), showing the spatial distribution of Cs, Pb, and Br.

| Pellet n = 1 | | |
|---|---|---|
| Element | Wt% | Atomic % |
| Br | 59.33 | 79.09 |
| Pb | 40.67 | 20.91 |
| Total | 100.00 | 100.00 |

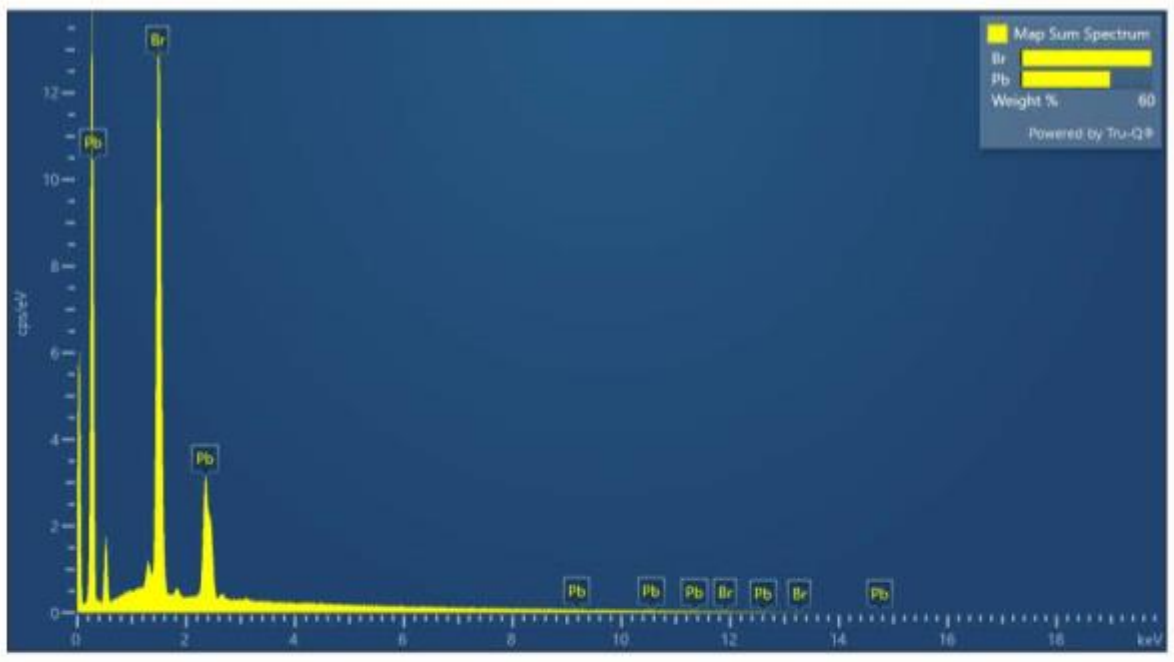

| Pellet n = 2 | | |
|---|---|---|
| Element | Wt% | Atomic % |
| Br | 51.17 | 70.53 |
| Cs | 11.83 | 9.80 |
| Pb | 37.00 | 19.67 |
| Total | 100.00 | 100.00 |

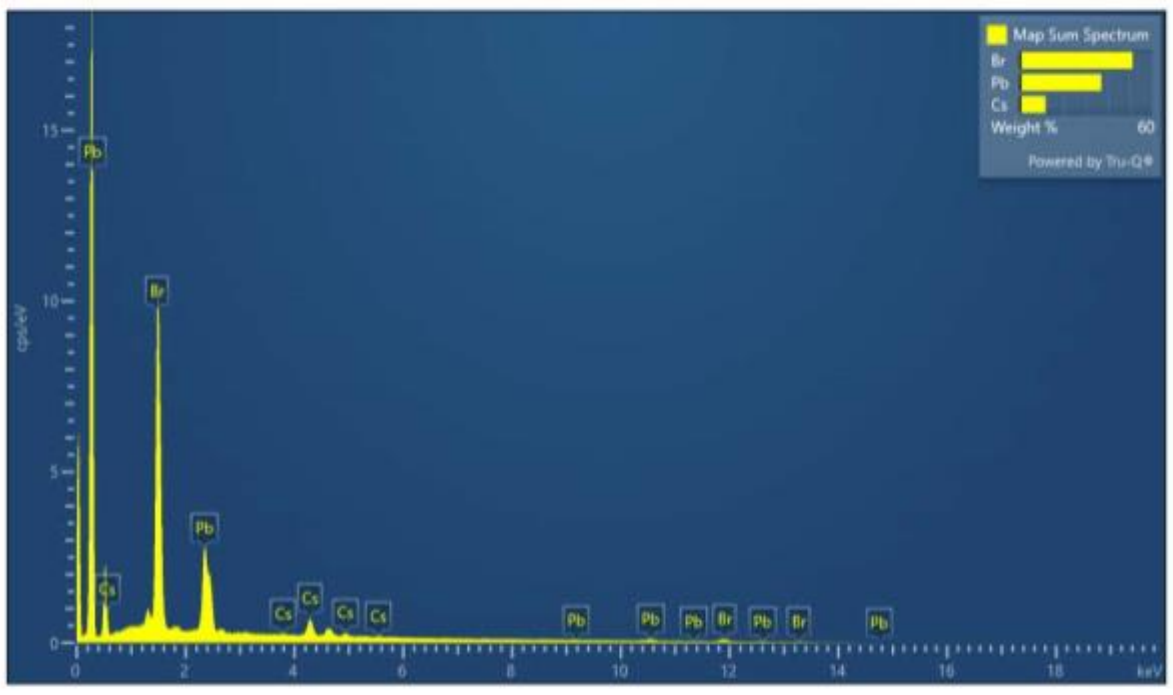

| Pellet n = 3 | | |
|---|---|---|
| Element | Wt% | Atomic % |
| Br | 47.11 | 66.30 |
| Cs | 16.45 | 13.92 |
| Pb | 36.44 | 19.78 |
| Total | 100.00 | 100.00 |

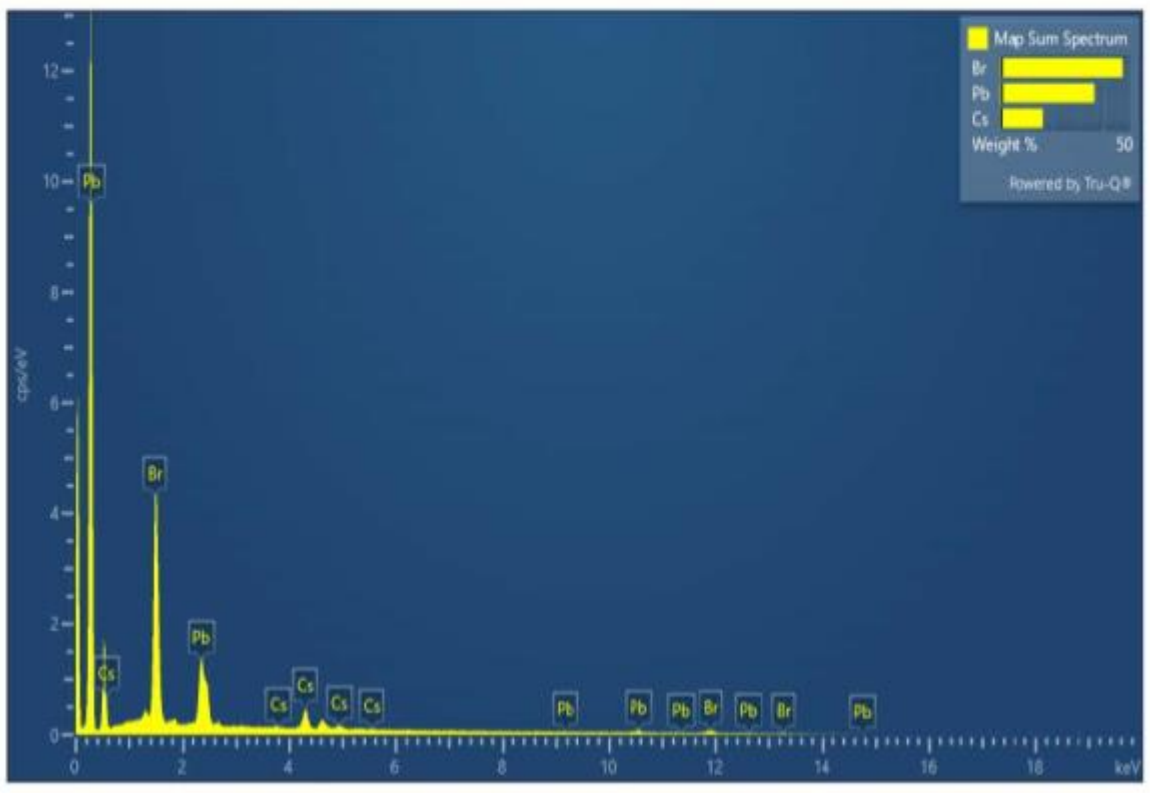

**Figure S18.** EDS spectra and corresponding elemental composition maps of the pellets pressed from the pure 2D perovskite microcrystals samples (*n* = 1–3), showing the spatial distribution of Cs, Pb, and Br.

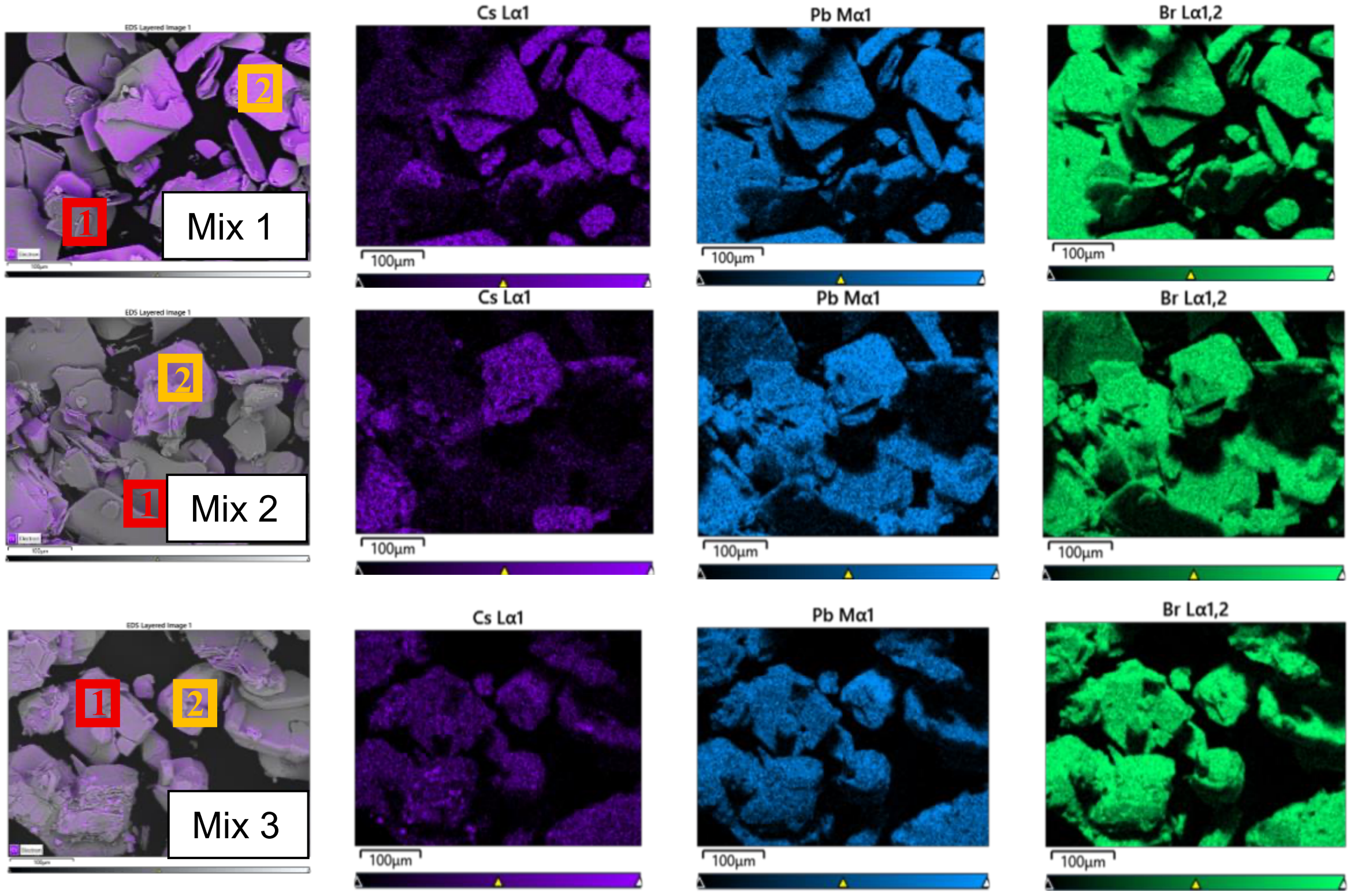

**Figure S19**. SEM images and corresponding EDS elemental maps of mixed quasi-2D perovskite microcrystal powders, showing the spatial distribution of Cs, Pb, and Br.

**Mix 1**

| Element | Wt% | Atomic % |
|---|---|---|
| | Spectrum 1 | |
| Pb | 41.81 | 21.70 |
| Br | 58.19 | 78.30 |
| Total | 100.00 | 100.00 |
| | Spectrum 2 | |
| Cs | 11.55 | 9.64 |
| Pb | 38.08 | 20.40 |
| Br | 50.37 | 69.96 |
| Total | 100.00 | 100.00 |

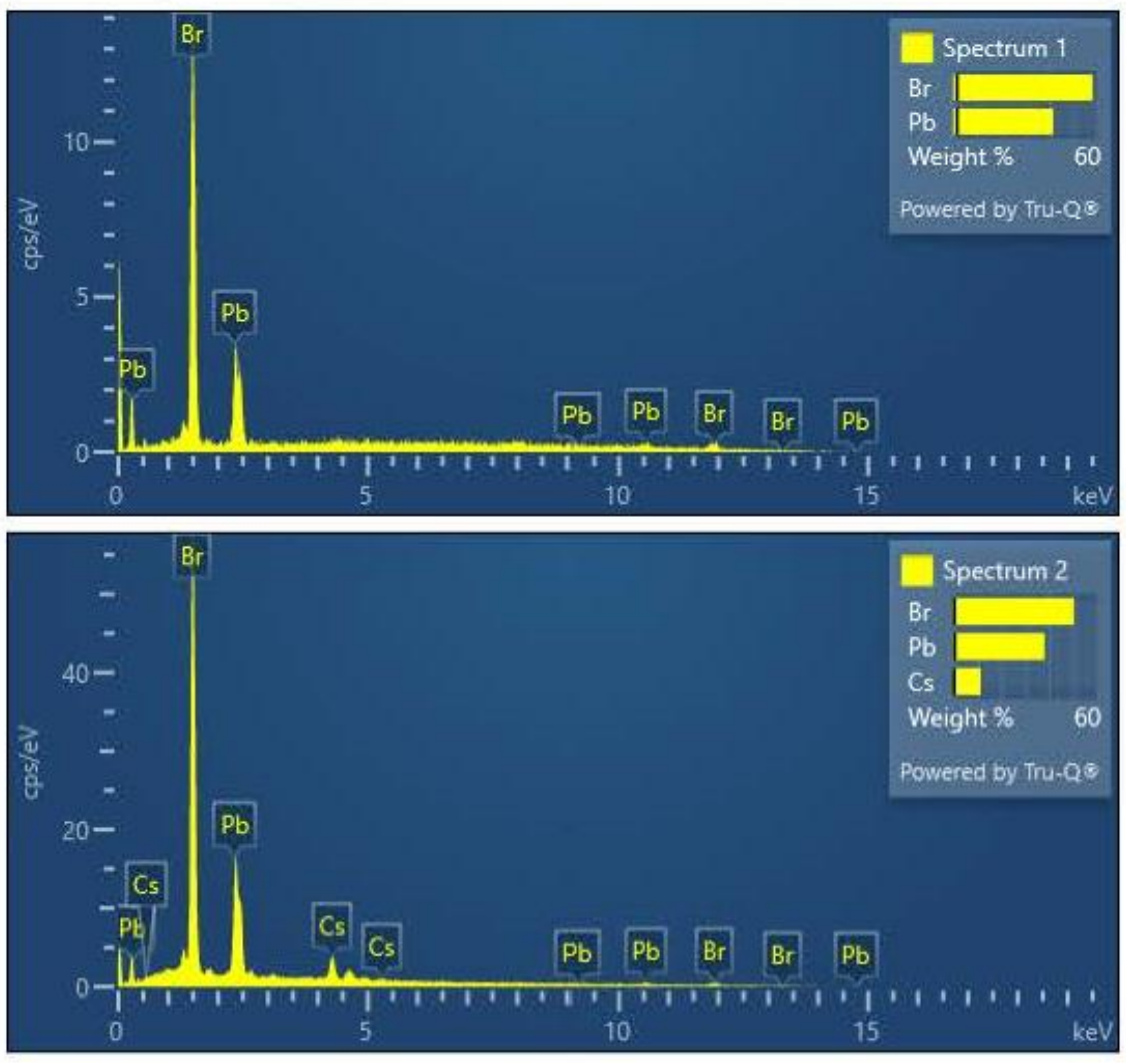

**Mix 2**

| Element | Wt% | Atomic % |
|---|---|---|
| | Spectrum 1 | |
| Pb | 40.60 | 20.86 |
| Br | 59.40 | 79.14 |
| Total | 100.00 | 100.00 |
| | Spectrum 2 | |
| Cs | 13.56 | 11.35 |
| Pb | 37.09 | 19.92 |
| Br | 49.35 | 68.73 |
| Total | 100.00 | 100.00 |

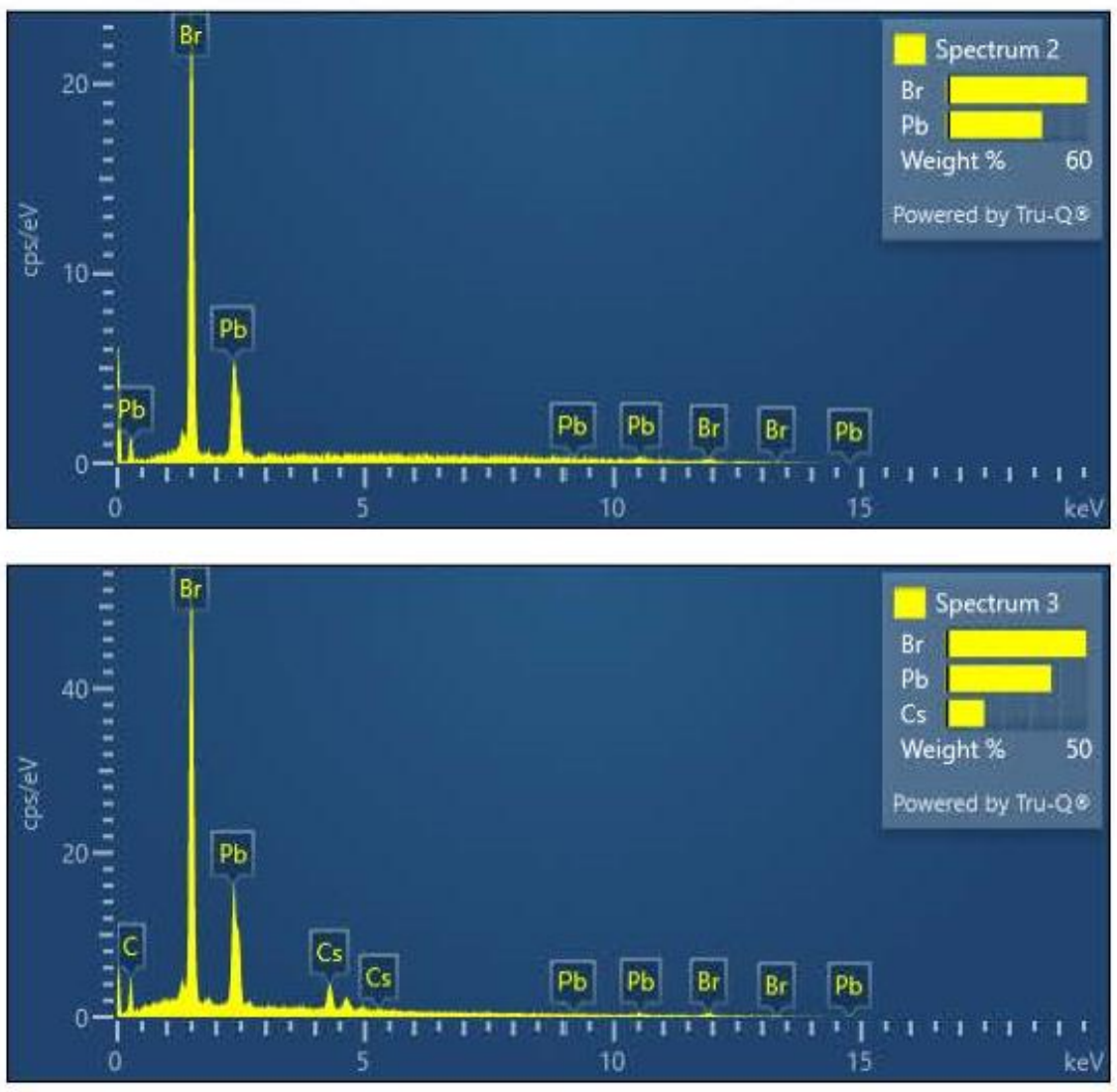

**Mix 3**

| Element | Wt% | Atomic % |
|---|---|---|
| | Spectrum 1 | |
| Cs | 13.31 | 11.44 |
| Pb | 40.35 | 22.26 |
| Br | 46.35 | 66.30 |
| Total | 100.00 | 100.00 |
| | Spectrum 2 | |
| Cs | 15.31 | 12.77 |
| Pb | 35.52 | 19.01 |
| Br | 49.17 | 68.22 |
| Total | 100.00 | 100.00 |

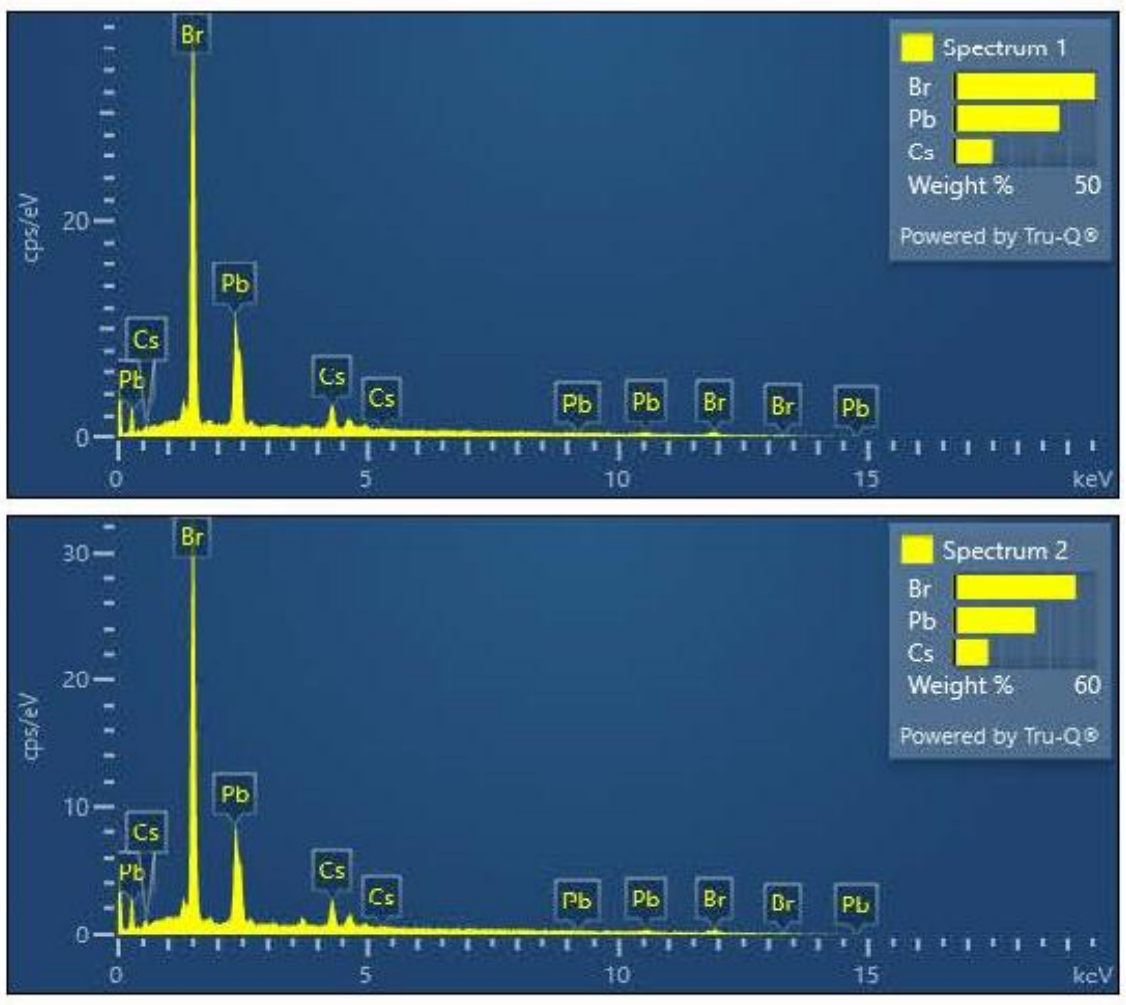

**Figure S20.** EDS spectra and corresponding elemental composition maps of mixed 2D perovskite microcrystal powders, confirming the presence of Pb, Cs, and Br.

From the responsivity R the detectivity D* can be calculated as:

$$D^* = \frac{R\sqrt{A}}{\sqrt{2eI_d}}$$

where $A$ is the active area of the device ($1 mm^2$), $e$ is the electronic charge, and $I_d$ is the dark current, derived from the region far below the band edge in the photocurrent spectra.

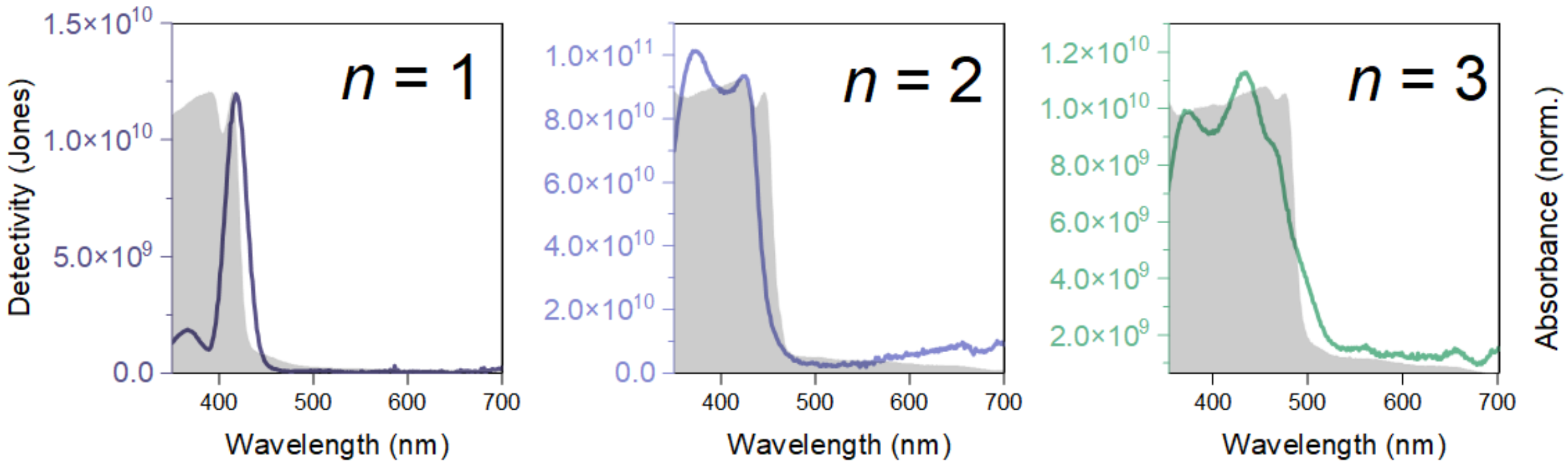

**Figure S21.** Detectivity spectra (colored) overlaid with the corresponding optical absorption spectra (gray) for pellet-based devices with $n$ = 1–3.